\documentclass[]{article}
\usepackage{authblk}

\title{Towards a (meta-)mathematical theory of consciousness: universal (mapping) properties of experience}

\author[1]{Steven Phillips}
\author[2, 3]{Naotsugu Tsuchiya}
\affil[1]{
	{\small National Institute of Advanced Industrial Science and Technology (AIST), Japan}
}
\affil[2]{
	{\small Laboratory of Qualia Structure, ATR Computational Neuroscience Laboratories, Japan}
}
\affil[3]{
	{\small Turner Institute for Brain and Mental Health, Monash University, Australia}
}

\date{}
	
\usepackage{amsmath, amssymb, amsthm, stmaryrd}
\usepackage{apacite}
\usepackage{enumerate}
\usepackage[a4paper, margin=25mm]{geometry}
\usepackage[utf8]{inputenc}
\usepackage{graphicx} \graphicspath{{./figures/}}
\usepackage{tikz-cd}
\usepackage{wasysym}

\DeclareMathOperator{\cod}{\mathit{cod}}
\DeclareMathOperator{\dom}{\mathit{dom}}

\DeclareMathOperator{\Hom}{Hom}
\DeclareMathOperator{\id}{\mathit{id}}
\DeclareMathOperator{\Lim}{Lim}

\DeclareMathOperator{\res}{\mathit{res}}

\newcommand\op{\mathrm{op}}
\newcommand\dotrightarrow{\dot\rightarrow}

\newcommand\invexcl{\rotatebox[origin=c]{180}{!}} 

\theoremstyle{definition}
\newtheorem{thm}{Theorem}

\newtheorem{defn}[thm]{Definition}

\newtheorem{exmp}[thm]{Example}
\newtheorem{exmps}[thm]{Examples}
\newtheorem{rem}[thm]{Remark}
\newtheorem{rems}[thm]{Remarks}

\begin{document}

\maketitle

\begin{abstract}
	Conscious experience permeates our daily lives, yet general consensus on a theory of consciousness remains elusive. In the face of such difficulty, an alternative strategy is to address a more general (meta-level) version of the problem for insights into the original problem at hand. Category theory was developed for this purpose, i.e. as an axiomatic (meta-)mathematical theory for comparison of mathematical structures, and so affords a (formally) formal approach towards a theory of consciousness. In this way, category theory is used for comparison with Information Integration Theory (IIT) as a supposed axiomatic theory of consciousness, which says that every conscious state involves six axiomatic properties: the IIT axioms for consciousness. All six axioms are shown to follow from the categorical notion of a universal mapping property: a unique-existence condition for all instances in the domain of interest. Accordingly, this categorical approach affords a formal basis for further development of a (meta-)mathematical theory of consciousness, whence the slogan, ``Consciousness is a universal property.''
\end{abstract}

\section{Introduction}\label{sect:introduction}

Conscious experience permeates our daily lives and the ``hard'' problem of consciousness is to explain why those subjective experiences feel in the way they do \cite{chalmers1995facing}. A generally accepted theory of consciousness remains elusive, however, despite millenia of study and numerous theoretical proposals \cite{doerig2021hard, kuhn2024landscape, seth2022theories}. In the face of such difficulty an alternative to a direct approach is to address more general aspects of the problem. For instance, the \textit{meta-problem of consciousness}, i.e. the problem of explaining why we think explaining consciousness is hard, and the hard problem are supposed to the connected: ``We can reasonably hope that a solution to the meta-problem will shed significant light on the hard problem.'' \cite<>[p.~8]{chalmers2018meta}. \textit{Category theory} \cite{leinster2014basic} is a kind of meta-mathematics that may facilitate doing ``ordinary'' mathematics by making precise the relationships between formal theories \cite{eilenberg1945general}. In this spirit, category theory is seen here as affording an alternative (meta-)theoretical approach towards consciousness (although the meta-problem of consciousness is not specifically targeted in the current work). 

The particular categorical approach taken here is motivated by several desiderata beyond category theory just as a form of meta-mathematics. Firstly, the quintessential challenge is to construct a theory of consciousness that explains subjective phenomena in a way that also transcends particular individuals. Category theory provides a formalism to address this challenge in terms of so-called \textit{universal mapping properties}, which has been used to provide a categorical account \cite{phillips2010categorial, phillips2016systematicity} for the systematicity of cognition \cite{fodor1988connectionism}. Intuitively, a universal mapping property is a property (in the form of a map) that is common to all entities in some class of entities under consideration. A universal mapping property acts like an a ``landmark'' that affords unique identification of every individual relative to that property. A familiar example is Coordinated Universal Time (UTC), which affords a ``unique'' reference to every local time zone relative to this temporal landmark (up to an equivalence). Category theory formalizes such situations in terms of entities, called \textit{objects} (e.g., time zones), and ``directed relations'' between objects, called \textit{maps}, \textit{morphisms} or \textit{arrows} (e.g., time zone differences). The basic idea is that subjective experience is composed of two components: a unique (subject-specific) component and a common (subject-invariant) component regarded as point of reference. In this way, category theory affords a formal treatment of subjective experience. 

Secondly, in lieu of direct access to the subjective states of other individuals, one may interrogate the internal structure and contents of those states by their (causal) relations to other states, for example, using report or no-report paradigms \cite<see>[for a discussion on the relative merits of these paradigms]{tsuchiya2015no}. This relational approach compares with category vs. set theory foundations for mathematics. In contrast to set-theoretic foundations, where constructions are given by sets whose elements are other sets (with the exception of the empty set, which has no elements), a category-theoretic foundation replaces set membership with a corresponding notion of arrow \cite<see>[]{leinster2014rethinking, lawvere2003sets}. This move avoids well-known conceptually unnatural consequences of some set-theoretic constructions, e.g., numbers are supposed to be founded on sets, $\pi$ is a number, so what are supposed to be the members of $\pi$? In category theory, numbers are more naturally seen in terms of their relationships to other numbers, i.e. as arrows, instead of elements \cite{maclane1986mathematics}. Likewise, subjective states can be characterized as arrows in relation to other subjective states \cite{tsuchiya2016using, tsuchiya2021relational}, though the contents of subjective experience need not be seen as exhaustively determined by such relationships \cite{negro2025three}.

And thirdly, category theory is an \textit{axiomatic} approach (whose examples and theorems derive from a small collection of basic assumptions---axioms) that yields insights by comparisons of theories \cite<see>[for perspectives]{kromer2007tool, marquis2009geometrical}. So, a natural place to start is with \textit{Integrated information theory} (IIT) as a supposed axiomatic theory of consciousness \cite{albantakis2023integrated, iit2024wiki, tononi2015integrated}. Although the axiomatic status of IIT has been called into question \cite{bayne2018axiomatic}, the intent is nonetheless clear: according to IIT, every state of consciousness is supposed to involve six (informally specified) properties that are the so-called ``axioms'' of subjective experience, which are operationalized by a corresponding list of ``postulates''. The axioms of IIT are presented next with a category theoretical interpretation pertaining to universal mapping properties, in the section that follows, and a discussion of this categorical account in the final section. The main point here is to show how this categorical approach affords a formal basis for further development of a (meta-)mathematical theory of consciousness.

\section{IIT as an axiomatic theory of consciousness}

IIT \cite{albantakis2023integrated, iit2024wiki, tononi2015integrated} is a prominent theory of consciousness, which asserts the existence of conscious experience from an intrinsic system of causal units. An attractive feature of IIT---for a formal (mathematical) theory of consciousness---is the characterization of phenomena by six axioms, or essential properties that are supposedly necessary and irrefutably true of every experience, briefly summarized in the next section.

Despite empirical support for aspects of IIT, this theory has been discounted as an axiomatic theory ``either because [IIT] fails to qualify as axiomatic or because it fails to impose a substantive constraint on a theory of consciousness.'' (\citeNP{bayne2018axiomatic}; see also \citeNP{cogitate2025adversarial, gomezmarin2025science, iitconcerned2025what}). Nonetheless, IIT continues to be strengthened axiomatically by upgrading the operational procedures defining the postulates. The most recent version of IIT \cite{albantakis2023integrated, iit2024wiki} affords more precise quantitative measures of how much conscious experience is present in a system and more precise qualitative assessments of what that experience feels like as the causal structure that is intrinsic to the physical system, thereby making a stronger case for IIT as an axiomatic theory.

Notwithstanding these advancements, there still remain two general problems for IIT as a \textit{theory} of consciousness. First, operational approaches are problematic when the core theoretical assumptions and principles are difficult to distinguish from auxiliary assumptions and conditions that are contingent on applications of the theory, especially in psychology \cite<see, e.g.,>[p.~1--4, in the context of cognitive psychology]{halford2014categorizing}---see also the \textit{systematicity} debate \cite{fodor1988connectionism, smolensky1988proper} regarding \textit{classicist} (symbol system) vs. \textit{connectionist} (neural network) theories of cognition, and the problem of \textit{ad hoc} auxiliary assumptions \cite{aizawa2003systematicity}. Second, computational procedures of the postulates are intractable for large systems, necessitating additional assumptions and heuristics for such situations \cite{albantakis2023integrated, iit2024wiki}, which in turn resonates with the first problem in blurring the distinction between core and contingent assumptions/principles.

As a redress to these problems towards a \textit{formally} axiomatic theory of consciousness, inspiration is taken from a parallel between the \textit{meta-problem of consciousness} \cite{chalmers2018meta, chalmers2020can} and \textit{category theory} \cite{leinster2014basic} as a kind of meta-mathematics \cite{eilenberg1945general} to be elaborated shortly. The meta-problem and the hard problem are supposed to be connected: ``We can reasonably hope that a solution to the meta-problem will shed significant light on the hard problem.'' \cite<>[p.~8]{chalmers2018meta}. Category theory is a meta-mathematical theory that may facilitate doing ``ordinary'' mathematics by making precise the relationships between formal structures \cite{eilenberg1945general}. In this way, category theory affords a perspicuous approach towards a (meta-)mathematical theory of consciousness, as the relationships between structures are typically well-defined---though the current work is not intended for the meta-problem. Towards this end, the axioms of IIT \cite{albantakis2023integrated, iit2024wiki} are regarded as a starting point to guide the choice of category theory concepts, as outlined in the rest of this extended introduction. This basic comparison between IIT axioms and category theory concepts motivates a (formally) formal, category theory treatment: the IIT axioms organize around three (meta-theoretic) principles, presented in the next section. A discussion of this category theoretical approach is given in the final section. Additional explanations are given in the appendices, including relevant mathematical definitions and examples (Appendix~A), and a comparison/contrast of terms such as axioms, properties and principles as used in regard to IIT and category theory (Appendix~B).

Some remarks before proceeding may also help guide the presentation style as befits a meta-theoretical approach. Category theory is not supposed to dictate what “the” theory of consciousness should be as relatively little is said in regard to the vast landscape of other consciousness theories \cite{kuhn2024landscape}. Category theory works well in mathematics and computer science where the base theories are typically well-established. However, the relationship between meta-theory and theory is analogous to the relationship between theory and data in that each side (level) is supposed to inform the other, as was suggested by the meta-problem and hard problem of consciousness in the first place. For this reason the presentation style is a back-and-forth between IIT on one hand and category theory on the other in trying to establish some structurally consistent correspondence, rather than a predominant focus on one as the driver of the other. IIT is taken as a starting point for its axiomatic foundations, which are amenable to a categorical treatment, not for any presumed preeminence or lack thereof as has been debated elsewhere \cite{tononi2025consciousness, iitconcerned2025what, gomezmarin2025science}. Note, however, that a categorical approach is not supposed to be one of just establishing correspondences between extant theories, as the base theories (which may also involve category theory) are still required to explain data. A theory is not only supposed to account for the relevant data, but also connect with the rest of the scientific enterprise. So category theory is seen as playing a crucial role in this regard.

\subsection{The axioms of IIT (revisited): essential properties}

IIT says that consciousness pertains to causal relations within a system \cite{albantakis2023integrated, iit2024wiki, tononi2015integrated}. The foundation of IIT is the \textit{axiom of existence}---zeroth axiom---and five axiomatic properties of every conscious experience \cite<as described by>[with quoted text and emphasis from there]{albantakis2023integrated}. 
\begin{enumerate}\setcounter{enumi}{-1}
	\item \textit{Existence}. ``Experience \textit{exists}: there is \textit{something}.''
	
	\item \textit{Intrinsicality}. ``Experience is \textit{intrinsic}: it exists for \textit{itself}.'' Experience is always first-person and independent of external observers.
	
	\item \textit{Information}. ``Experience is \textit{specific}: it is this \textit{one}.'' A moment of consciousness is informative because it picks out a specific experience from many other possible experiences for some event, not a generic abstraction.
	
	\item \textit{Integration}. ``Experience is \textit{unitary}: it is a \textit{whole}, irreducible to separate experiences.'' Experience cannot be split into independent consciousnesses, e.g., the experience of a \textit{red book} cannot be separated (disintegrated) into experiences of \textit{red} and \textit{book}.
	
	\item \textit{Exclusion}. ``Experience is \textit{definite}: it is this \textit{whole}.'' The phenomenon experienced is to the exclusion of all other phenomena that could have been experienced at that time.
	
	\item \textit{Composition}. ``Experience is \textit{structured}: it is composed of \textit{distinctions} and the \textit{relations} that bind them together, yielding a \textit{phenomenal structure} that feels \textit{the way it feels}.'' A moment of experience is composed of various aspects of experience and their relationships. 
	
\end{enumerate}
Axioms 1--5 are taken to be the \textit{essential} properties of every conscious experience. 

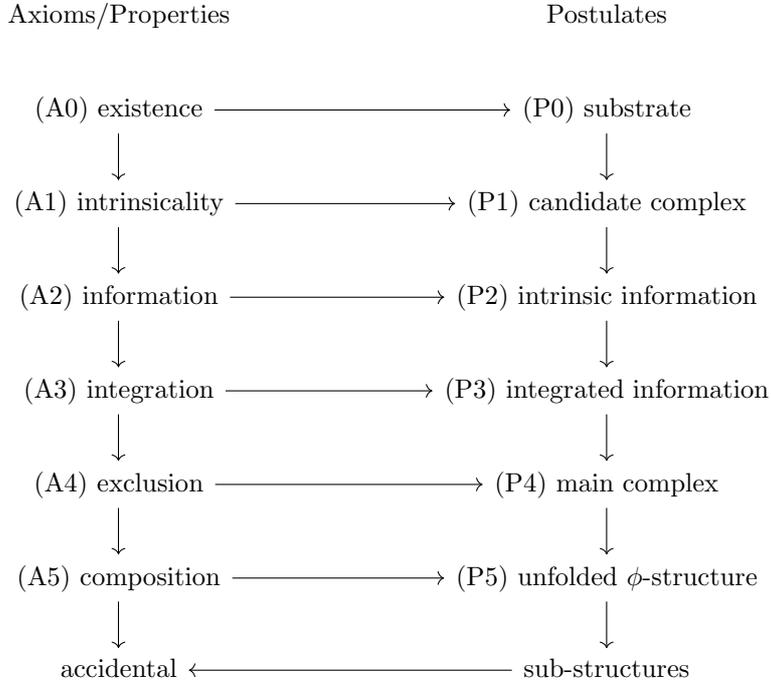
\begin{figure}[ht]
	\begin{equation*}
		\begin{tikzcd}
			\mbox{Axioms/Properties} &&& \mbox{Postulates} \\
			\mbox{(A0) existence} \arrow[rrr] \arrow[d] &&& \mbox{(P0) substrate} \arrow[d] \\
			\mbox{(A1) intrinsicality} \arrow[rrr] \arrow[d] &&& \mbox{(P1) candidate complex} \arrow[d] \\
			\mbox{(A2) information} \arrow[rrr] \arrow[d] &&& \mbox{(P2) intrinsic information} \arrow[d] \\
			\mbox{(A3) integration} \arrow[rrr] \arrow[d] &&& \mbox{(P3) integrated information} \arrow[d] \\
			\mbox{(A4) exclusion} \arrow[rrr] \arrow[d] &&& \mbox{(P4) main complex} \arrow[d] \\
			\mbox{(A5) composition} \arrow[rrr] \arrow[d] &&& \mbox{(P5) unfolded $\phi$-structure} \arrow[d] \\
			\mbox{accidental} &&& \mbox{sub-structures} \arrow[lll]
		\end{tikzcd}
	\end{equation*}	
	\caption{IIT axioms, properties and postulates.}\label{fig:iit axioms}
\end{figure}

All six axioms are operationalized by translation to corresponding postulates. This translation is in order of the axioms and so starts with the corresponding \textit{existence postulate}, i.e. by specifying a so-called \textit{substrate model} of the system that is operationalized as a \textit{transition probability matrix} (TPM) encoding the substrate of consciousness \cite{albantakis2023integrated, iit2024wiki}. Applying the other five postulates to this model unfolds a \textit{cause-effect} ($\phi$-)structure with possible \textit{accidental} (additional) properties for the application of IIT at hand \cite<see>[for further details and delineations of IIT concepts]{grasso2024glossary}. The relationships between IIT axioms, properties and postulates are shown in Figure~\ref{fig:iit axioms}. For example, the ``extendedness'' of phenomenal space is derived from applying the postulates to a suitable substract model, which has a formal correspondence to the axioms of a topological space \cite{haun2019does, haun2024unfathomable}, as summarized next.

\subsubsection{Phenomenology of space (example): accidental properties}

An example illustrating the relationships between the axioms, (essential and accidental) properties and postulates is the phenomenology of spatial experience \cite{haun2019does, haun2024unfathomable}, whereby the night sky is supposed to involve:
\begin{itemize}
	\item an ``extended canvas'' on which one can distinguish
	\item ``spots'' (regions of the canvas) that are 
	\item ``extended'' (to constitute an arrangement of spots) in the sense of being:
	\begin{itemize}
		\item ``connected'' to other spots as their overlaps that are also spots,
		\item ``fused'' with other spots as their amalgamations that are also spots and
		\item ``included'' in spots as one spot being inside another spot, or equal to itself,
	\end{itemize}
	\item along with other ``sub-structures'' of experience, e.g., location and distance.
\end{itemize}
In axiomatic terms: (0) phenomenal space exists, (1) as experienced by/for the subject, (2) in picking out one specific configuration of spots and their relations, (3) as a single collection of spots, (4) excluding other possible collections of spots, for that moment of experience, and (5) composed of distinct spots and their connections, fusions and inclusions, among other possible accidental (sub-structural) properties. These phenomenal properties are determined from the postulates by first specifying, in this case, a grid-like substrate (encoded as a TPM) and applying the computational procedures, as defined by the postulates, accordingly. In doing so, the unfolded (cause-effect) $\phi$-structure corresponds to the composition property, whose sub-structures yield the accidental properties of phenomenological space, including extendedness, connectedness and so on \cite{haun2019does, haun2024unfathomable}.

This (informal) characterization of the properties for phenomenal space has a formal counterpart, in set-theoretic terms, by supposing that the extended canvas is a set of points and spots are subsets of that canvas, so being connected, fused and included corresponds to set intersection, union and inclusion, respectively. If spots are supposed to be \textit{open sets}, in the topological sense, then the phenomenal property of extendedness corresponds to the compositional structure of a topological space \cite{haun2019does}. A \textit{topological space} (definition~\ref{defn:topological space}) captures the notion of proximity (remark~\ref{rem:topological space}) as a collection of open sets, which is also an \textit{ordered set} (definition~\ref{defn:poset}) where the open sets are (partially) ordered by inclusion (remark~\ref{rem:poset}). A formal correspondence to the IIT axioms follows: (0) the existence of a topological space $(X, T)$, (1) as internal to some larger set, pertaining to the subject, (2) a specific topology $T$ on $X$, among many other possible topologies on $X$, (3) the integrated set $T$ of open sets, (4) to the exclusion of other possible topologies on $X$ and (5) composed of open sets $U, V$ of $T$ and inclusion relations $V \subseteq U$. The phenomenological feeling of extendedness is then formally the topology whose ``distinctions and relations that bind them'' are the open sets (which includes intersections and unions) and inclusion relations. 

\subsection{Counterpoint: subjective-first vs. objective-first}

The quintessential view driving IIT is a \textit{subjective-first} approach to phenomenology, in contradistinction to the \textit{objective-first} approach that prevails elsewhere in the study of consciousness and in the other sciences \cite{albantakis2023integrated, iit2024wiki, tononi2015integrated}---cf. egocentric vs. allocentric points of reference, or first-person vs. third-point points of view. So, instead of starting with the axioms for a topological space from which the phenomenology of space may be derived, IIT starts with the phenomenal properties of space, encoded by a system of units as a subject, to derive the phenomenal (topological) quality of space, by unfolding the system's intrinsic causal relations. Likewise, this subjective-first approach has been taken to give an IIT account for the phenomenal flow of time as directed space \cite{comolatti2024time}. 

The axioms and (essential or accidental) properties for different phenomenal experiences may be akin to, but need not be the same as the axioms and properties of mathematical structures. To wit, ``sub-structures'' of phenomenal space, such as phenomenal distance, suggest a formal correspondence to a \textit{metric space} (definition~\ref{defn:metric space}). Every metric space is a topological space---justifying the notion of phenomenal distance as a sub-structure of phenomenal space---but a topological space need not be a metric space (remark~\ref{rem:metric space}). From a formal perspective, this asymmetry comports with the IIT view that the five properties (Axioms~1--5) are necessary (essential), but not necessarily sufficient for some phenomenal experiences (accidental properties)---together, the axioms for topological spaces are necessary, but not sufficient for metric spaces; conversely, the axioms for metric spaces are (conjointly) sufficient, but not necessary for topological spaces. A similar situation arises when trying to formalize phenomenal space as a \textit{measurable space} (definition~\ref{defn:measurable space}) compared to topological space (remark~\ref{rem:measurable space}) to capture other properties by a \textit{measure space} (definition~\ref{defn:measure space}) for, say, phenomenal vividness or certainty (remark~\ref{rem:probability space}). Axioms in the formal mathematical (definitional) sense are usually seen as necessary and sufficient conditions for something to be a structure of some kind. Each instance of a structure may have additional properties that are derived from the axioms as theorems, or are particular to some instances. In group theory, for example, an axiom (or essential property) of every group is having an identity element, but the uniqueness of the identity element is a theorem (or accidental property) derived from the axioms; the existence of a unique group homomorphism to/from every group is a property that only pertains to one-element groups. Computational methods (or models) determining such properties are more akin to the postulates of IIT. 

Although mathematics may be regarded as objective, the development of mathematics is also seen as arising from the embodiment of cognition \cite{lakoff2000mathematics} and from the formalization of conceptual experience \cite{lawvere2009conceptual, maclane1986mathematics}. And so, mathematical experience need not be entirely divorced from the subjective-first perspective. The approach taken here may be seen as conciliatory with the view that at some point one's subjective experiences are to be reconciled with the experiences of others. 

\subsection{Towards formal axioms for consciousness}

The meta-level approach (alluded to earlier) is to view the properties of phenomenal experience from meta-theoretical considerations---cf. approaching the hard problem of consciousness via a solution to the meta-problem \cite{chalmers2018meta, chalmers2020can}. Category theory is a natural meta-mathematical choice as topological spaces are instances of \textit{categories}, in the category theory sense, whereby intersection, union and inclusion pertain to particular kinds of category theoretical structures. Furthermore, the relations between different mathematical structures, such as topological and metric spaces, and their (meta-level) categories are generally well-understood in category theory terms. The approach taken here is to educe category (meta-)theoretical principles to obtain axioms for phenomenal experience. These principles and their relationships to the axioms of IIT are laid out next.

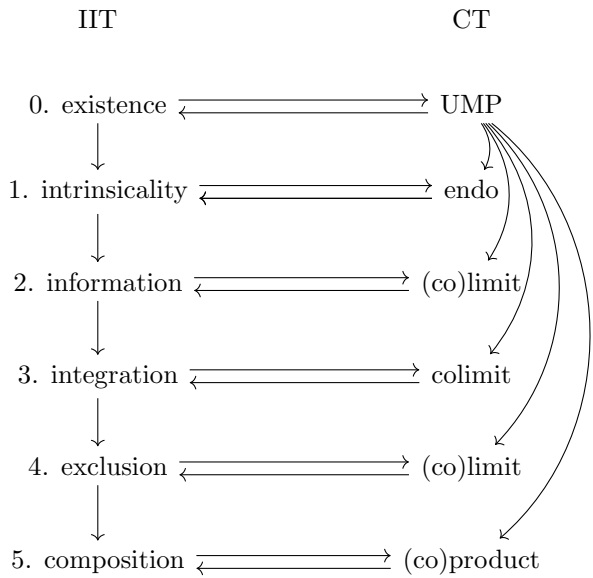
\begin{figure}[ht]
	\begin{equation*}
		\begin{tikzcd}
			\mbox{IIT} &&& \mbox{CT} \\
			\mbox{0. existence} \arrow[rrr, shift left] \arrow[d] &&& \mbox{UMP} \arrow[lll, shift left] \arrow[d, bend left=30] \arrow[dd, bend left=35] \arrow[ddd, bend left=40] \arrow[dddd, bend left=45] \arrow[ddddd, bend left=50] \\
			\mbox{1. intrinsicality} \arrow[rrr, shift left] \arrow[d] &&& \mbox{endo} \arrow[lll, shift left] \arrow[lll, shift left] \\
			\mbox{2. information} \arrow[rrr, shift left] \arrow[d] &&& \mbox{(co)limit} \arrow[lll, shift left] \\
			\mbox{3. integration} \arrow[rrr, shift left] \arrow[d] &&& \mbox{colimit} \arrow[lll, shift left] \\
			\mbox{4. exclusion} \arrow[rrr, shift left] \arrow[d] &&& \mbox{(co)limit} \arrow[lll, shift left] \\
			\mbox{5. composition} \arrow[rrr, shift left] &&& \mbox{(co)product} \arrow[lll, shift left] \\
		\end{tikzcd}
	\end{equation*}	
	\caption{A comparison of Integrated Information Theory (IIT) axioms/postulates in relation to category theory (CT) structures that satisfy a universal mapping property (UMP) of a particular kind. The left downward arrows indicated the computational order for the postulates. The right downward arrows indicate the particular kind of UMP. Horizontal arrows are correspondences---a rightward arrow is a formalization of the axiom as a UMP of a particular kind. The CT properties are not ordered.}\label{fig:iit ct}
\end{figure}

A core category theory principle is characterization of a mathematical structure by a so-called \textit{universal mapping property} (UMP), i.e. a \textit{unique-existence} condition satisfied by all instances of the structure \cite{leinster2014basic, maclane1998categories}. For some intuition, the largest (maximum) element in an ordered set satisfies a UMP: every element in that set is smaller than or equal to the largest element. For instance, infinity satisfies a UMP in being the largest number in the set of natural numbers extended to include infinity. Likewise, the smallest (minimum) element satisfies a UMP: every element in that set is larger than or equal to the smallest element. For instance, zero satisfies a UMP in being the smallest natural number. IIT employs two principles to determine cause-effect structure: \textit{maximal/minimal principles of existence}, i.e. in very simple terms, the maximum collection or minimum partitioning of units that make or take a difference \cite{albantakis2023integrated, iit2024wiki}. So, these principles for determining cause-effect structure pertain to UMPs. For instance, the accidental properties of connection and fusion for phenomenal space, derived by the postulates, are the UMPs for set intersection and union, respectively. The meta-theoretical approach taken here is to formalize IIT axioms in relation to UMPs. A summary of relationships between IIT and UMPs is given in Figure~\ref{fig:iit ct} as a reference guide.

\section{Category theoretical axioms and principles}\label{sect:category axioms}

The move from theory (IIT) to meta-theory (category theory) faces a number of choices as there may be many possible avenues of abstraction. ``Put differently, good general theory does not search for the maximum generality, but for the right generality.'' \cite<>[p.~108]{maclane1998categories}. Moreover, the most direct route is not always the shortest route, as exemplified with the invention of category theory: `As Eilenberg-Mac Lane first observed, ``category'' has been defined in order to be able to define ``functor'' and ``functor'' has been defined in order to be able to define ``natural transformation''.' \cite<>[p.~18]{maclane1998categories}. What determines the ``right'' generality or the ``shortest'' route depends on the task at hand. So, two preliminary remarks are in order to frame the approach and choices that follow: in particular, a predominant focus on the composition axiom ahead of the other axioms.

First, recall that IIT starts by asserting ``Experience exists'' (zeroth axiom---existence) and then by asserting ``it exists for itself'' (first axiom---intrinsicality) defined operationally as ``cause-effect power'' encoded as a matrix of transition probabilities between units (zeroth postulate---substrate) to distinguish units in the complex, which directly contribute to experience, from units outside the complex, including those that provide inputs to or outputs from the complex (first postulate---candidate complex). The view driving IIT is that the quality (feeling) of phenomenal experience is intrinsic causal structure, i.e. the \textit{inform}ational relations between units that must ``take and make a difference'', where to \textit{inform} is taken in the etymological sense, i.e. ``to give character or \textit{essence} to; to be the characteristic \textit{quality} of'' \cite<>[emphasis here]{merriamNDinform}, not in the \textit{information theory} \cite{shannon1948mathematical} sense of communication \cite{zaeemzadeh2024shannon}. Category theory affords a particular interpretation of causal relations in terms of \textit{arrows} (also called \textit{morphisms}) between \textit{objects} in some \textit{category} (to be defined shortly) in supposing that ``A causes B'' corresponds to an arrow from an object $A$ to an object $B$, written $A \rightarrow B$, in some collection of such causal relations of a system as a category \cite{otsuka2024process}. (Objects may correspond to simple units or more complex entities with their own internal structure.) Furthermore, to \textit{in-form} is to send \textit{in}to $B$ the \textit{form} (shape) of $A$, suggesting a particular kind of arrow. The zeroth postulate, then, corresponds to specifying a category as a substrate model and the first postulate as distinguishing candidate parts of the category that are extrinsic versus intrinsic to conscious experience.

Second, although interpreting causal relations as arrows is straightforward, there are still many choices of categories and ways to distinguish arrows for an axiomatic theory of phenomenal experience. So, these two axioms/postulates are not particularly informative in this regard. On the other hand, the phenomenal space example showed how the quality of experience pertains to the composition axiom/postulate as the structure of a topological space. Topological spaces are straightforwardly categories and the intersections and unions of sets are simple instances of structures satisfying UMPs (as already mentioned) in a category whose ``units'' are sets related by inclusions. This more informative connection between IIT and category theory suggests another way forward, towards a meta-mathematical theory, by taking a categorical perspective on IIT from the example of phenomenal space \cite{haun2019does} with an initial focus on the composition axiom and the maximum/minimum existence principles in terms of UMPs. The next section introduces the basic category theory needed to pursue this approach in the sections that follow.

\subsection{Categorical axioms}

A \textit{category} (definition~\ref{defn:category}) is similar in structure to a \textit{(directed) graph} (remark~\ref{rem:graph}) in consisting of a collection of entities, called \textit{objects} (cf. nodes), and a collection of ``directed relations'' between objects, called \textit{arrows}, \textit{maps}, or \textit{morphisms} (cf. edges). However, categories have additional structure in the form of an operation on pairs of arrows, called \textit{composition}, and a collection of ``reflexive relations'', called \textit{identity arrows}, or simply \textit{identities} that correspond to additional edges for paths and loops (self-directed edges), respectively, that graphs need not have. This axiomatization of categories and graphs highlights another asymmetry: every category corresponds to a (directed) graph, but not every graph corresponds to a category (cf. metric and topological spaces, mentioned earlier). Archetypal examples of categories are the category of sets and functions, $\mathbf{Set}$ (example~\ref{exmp:sets}), and the category of sets and inclusions, $\mathbf{Set}^\subseteq$ (example~\ref{exmp:inclusions}), which is a \textit{subcategory} of $\mathbf{Set}$ (remark~\ref{rem:subcategory})---cf. subset---exemplifying a formal notion of sub-structure in a categorical sense. Topological spaces are categories (example~\ref{exmp:topological space}), as too are ordered sets (example~\ref{exmp:poset}). 

Intersections and unions of open sets in a topological space are instances of \textit{universal constructions}, a generic term for a variety of constructions having a \textit{universal mapping property} whose definitions follow a familiar pattern that specifies a unique-existence condition (remarks~\ref{rem:ump})---intuitively, every construction in the domain of interest is related by this property \cite<for more intuition see>[p.~1--7]{leinster2014basic}. Certain categories have certain kinds of universal constructions. Two important examples are \textit{categorical product} (definition~\ref{defn:product}), which is Cartesian product in $\mathbf{Set}$ (example~\ref{exmp:cartesian product}) and set intersection in $\mathbf{Set}^\subseteq$ (example~\ref{exmp:intersection}), and \textit{categorical coproduct} (definition~\ref{defn:coproduct}), which is disjoint union in $\mathbf{Set}$ (example~\ref{exmp:disjoint union}) and set union in $\mathbf{Set}^\subseteq$ (example~\ref{exmp:set union}). So, intersections and unions are products and coproducts in topological spaces. Two other important examples of universal constructions are \textit{terminal object} (definition~\ref{defn:terminal}), which is any singleton set in $\mathbf{Set}$ (example~\ref{exmp:singleton set}) and the total set in a topological space (example~\ref{exmp:total set}), and \textit{initial object} (definition~\ref{defn:initial}), which is the empty set in $\mathbf{Set}$ and in any topological space as a category (example~\ref{exmp:empty set}). All universal constructions are unique up to an isomorphism, hence usually referred to as ``the'' universal construction (remark~\ref{rem:uniqueness}), so a category may have more than one instance of a particular universal construction (example~\ref{exmp:equivalence}).

The translation from set-theoretic to category-theoretic versions of the axioms alludes to two (meta-theoretic) principles driving category theoretical approaches: (1) universality---identification of structures via universal mapping properties, e.g., products and terminal objects, and (2) duality---arrow reversal, e.g., coproducts and initial objects, also called \textit{coterminal objects}. These principles are intertwined: e.g., product and coproduct are duals (remarks~\ref{rem:duality}) and every arrow has just one (reversal of) direction, which are manifest in a categorical (re)formulation of the axioms for phenomenal space, canvassed next.

\subsubsection{Categorical axioms for phenomenal space}

Having recast topological concepts categorically, a correspondence between the six IIT axioms and formal, category theory concepts with regard to phenomenal space is straightforward (as a first-pass):
\begin{enumerate}\setcounter{enumi}{-1}
	\item existence: as the existence of a topological space as a category, $(X, T)$,
	
	\item intrinsicality: as a category $(X, T)$ that is internal to some larger category,
	
	\item information: as a specific category of open sets and inclusions, $(X, T)$, not some abstract category,
	
	\item integration: as the set (topology $T$) of open sets of $X$, 
	
	\item exclusion: as excluding other possible topologies $T'$ on $X$ and
	
	\item composition: as composed of objects (open sets) and arrows (inclusions) as distinctions and binding relations in a category with products (intersections) and coproducts (unions) yielding the extended feeling of space.
	
\end{enumerate}
So, the extendedness of the night sky is obtained by the distinctive spots (objects) that are connected (products), fused (coproducts) and related by inclusion (arrows). 

\subsubsection{Interim assessment of the axioms}

This formal (categorical) version of the axioms for phenomenal space makes apparent that most of the ``work'' is shouldered by the composition axiom. Like the set-theoretic version, the other axioms appear to be (more or less) different ways of appealing to the existence of some particular topological space, as a category. The category theory approach makes this situation even more apparent, since a collection of structures and structure-preserving (homo)morphisms is typically a category (remark~\ref{rem:structure categories}), such as the collection of topological spaces, $\mathbf{Top}$, whose arrows are the continuous functions. Every topological space, now as an object in $\mathbf{Top}$, is specific in being that object and nothing else (information) and cannot be reduced without becoming another object (integration). 

Clearly, however, more is intended from the IIT axioms, as not every topological space is supposed to correspond to a phenomenal experience. This additional work is taken up by the postulates via measures of causal interaction and guiding principles of maximization/minimization to identify the quality of experience by unfolding the cause-effect structure and thereby deriving the phenomenology of space as a particular topological space \cite{haun2019does}. However, the postulates are not formally related to the axioms, so IIT axioms do no (formal) work in this regard.

Category theory, by contrast, axiomatizes notions of ``best'' structure (e.g., intersection and union) in terms of universal mapping properties, as mentioned. The principles of maximal and minimal existence suggest some kind of universal mapping property, since the maximum and minimum elements of an ordered set (if such elements exist) as a category are the terminal and initial objects, respectively. Such axiomatization also extends to the (meta-)level of categories whose objects are other categories, e.g., in a category of topological spaces with the discrete space as the initial object (as another category). So, the move to such meta-theoretical principles is taken up next.

\subsection{Two categorical (meta-theoretical) principles}

Set intersection and union are determined from two general (meta-theoretical) principles that pervade applications of category theory to mathematics and elsewhere, either structurally or computationally: (1) \textit{universality}---construction from universal mapping properties---and (2) \textit{duality}---construction from arrow reversal. These principles are intimately related through the category acting as the frame of reference in demarcating \textit{internal} vs. \textit{external} structure, i.e. the objects and arrows that ``live'' inside vs. outside the category, and associated points of view. The internal perspective has been given to this point as the objects and arrows in some category. Universal mapping properties are given with respect to a category (internal view), or with respect to relations between categories (external view), which is akin to the IIT notion of cause-effect relations within and between partitions of unit \cite{iit2024wiki}. Note that external in one context (category) may be internal in another context, as categories may be objects in larger categories; likewise for universal constructions as exemplified by the following views. An advantage of taking a categorical approach is that such changes in perspectives and their relationships are well-defined and understood in category theory.

\paragraph{Internal view}
Recall that in the category of sets and inclusions, $\mathbf{Set}^\subseteq$, the intersection of sets $A$ and $B$ is a universal construction, which affords a simple example to outline the principle of construction from a universal mapping property. This construction requires some way of picking out $A$ and $B$ and a candidate set for intersection, i.e. an index-like map relating categories, since $A$ and $B$ are internal to $\mathbf{Set}^\subseteq$. A map between categories is a \textit{functor} (definition~\ref{defn:functor})---cf. graph homomorphism (remark~\ref{rem:functor})---and the categorical analog of an indexed set is a \textit{diagram} (example~\ref{exmp:diagram}), which is a functor that picks out a subcollection of objects and arrows---cf. indexed set (remark~\ref{rem:diagram}). Comparison of functors, or a map between functors is a \textit{natural transformation} (definition~\ref{defn:natural transformation}, examples~\ref{exmp:natural transformations}). (Functors and natural transformations also constitute a category, remark~\ref{rem:functor category}). In particular, the intersection of $A$ and $B$ is a set that is contained in both $A$ and $B$, i.e. a set that contains no fewer elements than every element that appears in $A$ and appears in $B$, and no more than those elements. The comparison is a \textit{cone} (definition~\ref{defn:cone}) that, for this situation, consists of a set $Z$ (remark~\ref{rem:cone}) as a candidate for intersection and the inclusions $Z \subseteq A$ and $Z \subseteq B$ (example~\ref{exmp:inclusion of intersection}). The comparison with other candidates is a cone homomorphism (definition~\ref{defn:cone homomorphism}), which is another inclusion in this instance (example~\ref{exmp:intersection subset}). The ``best'' candidate is the \textit{limit} (definition~\ref{defn:limit}), which is set intersection here (example~\ref{exmp:intersection limit}). (The terminal object is also a limit, remark~\ref{rem:terminal}.) The limit is the terminal object in a category of such cones (remark~\ref{rem:limit}), hence intersection is a universal construction. Limits as universal cones exemplify the internal view of universal construction. 

\paragraph{External view}
An (equivalent) external view of limit is defined in the form of \textit{universal morphism} (definition~\ref{defn:universal morphism}). By this view, a limit is a universal morphism from a \textit{diagonal functor} to a diagram, (remark~\ref{rem:limit as univeral morphism}), hence intersection is given by the universal morphism to the pair of sets $(A, B)$ as a diagram (example~\ref{exmp:product as universal morphism}). This distinction between internal vs. external is relative to a category. For instance, the external view of a limit as a universal morphism also has an internal view relative to a \textit{comma category} (definition~\ref{defn:comma category}), as every universal morphism is either a terminal or initial object in the appropriate comma category (remark~\ref{rem:universal morphism}). This internalization of external relations is important for the intrinsicality axiom, since phenomenal experience is supposed to be for the subject, not some external observer, which accommodates interactions within and between subsystems of a system. All universal constructions, including coproducts (example~\ref{exmp:coproduct-dual-universal}), dualize in these ways (remark~\ref{rem:dualize}). Diagrams (and arrows, generally) are formally/conceptually ``generalized elements'', hence the formal connection between diagrams and causal relations in the informational sense of IIT (remarks~\ref{rem:inform}), presaged in the lead for this section.

\subsubsection{Categorical axioms for phenomenal space (revisited)}

With the universality and duality principles at hand, the axioms for phenomenal space are revisited. (There are several, closely related notions of duality in category theory: as a definition or construction in the opposite category, in this section, or in regard to a pair of opposing arrows, as used later. The internal vs. external distinction may also be seen as an instance of the latter.) Observe that the composition axiom in regard to phenomenal space is essentially a claim about certain kinds of universal constructions, specifically, products and coproducts, and that the total set is also a universal construction, specifically the terminal object in that topology, which is internal to the category of topological spaces, $\mathbf{Top}$. Axioms~1--5 pertain to a universal construction of some kind. A construction is universal in the ``ambient'' category, i.e. the category within which the construction satisfies a universal mapping property. In particular, the total set $X$ is universal with regard to the given topology as the ambient category. The topology is universal with regard to a particular collection of (basis) subspaces as the ambient category, i.e. a subset of open sets from which the topology $T$ is recovered as the terminal object in a collection of such subspaces. So, in regard to the axioms, there is a topological space (existence) in a collection of spaces (intrinsicality) that is determined by a universal mapping property among those subspaces (information) as one topology (integration) which excludes other possible topologies that do not satisfy the unique-existence condition for universality (exclusion) consisting of products and coproducts of open sets and their inclusions (composition).

\subsubsection{Interim (re)assessment of the axioms}

The unique-existence condition formally underwrites interpretations of maximal and minimal existence principles. For comparison, the intersection of $A$ and $B$ is the set $P = A \cap B$ and two inclusions, $P \subseteq A$ and $P \subseteq B$, recovering every element that is in both $A$ and $B$. Any candidate set $P'$ with more elements than $P$, i.e. $P \subset P'$, is sufficient, but not necessary for recovery of those elements ($P'$ satisfies existence, but not uniqueness). So, $P$ is the least such set---cf. principle of minimal existence, i.e. ``nothing exists more than the least it exists'' \cite{grasso2024glossary}. Conversely, any candidate set $P''$ with less elements than $P$, i.e. $P'' \subset P$, is necessary, but not sufficient ($P''$ satisfies uniqueness, but not existence). So, $P$ is the most such set---cf. principle of maximal existence, i.e. ``what exists is what exists the most'' \cite{grasso2024glossary}. The same situation applies dually to union as a coproduct. In both situations, the maximum and minimum principles correspond to the two sides of the same unique-existence condition. The information and integration axioms, which depend on applying the maximum and minimum principles of the postulates, would appear to be derivable from UMPs. Likewise, the exclusion axiom also appears to derive from a UMP as the maximum complex. 

The phenomenal space example, however, is just one application used to develop the (meta-)theory. Other examples may exercise the other axioms. In particular, more than one experience may be associated with a given space, e.g., the sky at night vs. daytime. This situation motivates a third principle which was tacit in the previous analysis, whereby constructions are localized to a particular category, that reveals how these IIT axioms may be independent. This locality principle is made explicit by another kind of construction, next.

\subsection{Third principle: locality}

Space may include other phenomena beyond extendedness, which may also vary in relation to proximity and from moment to moment. The notion of properties localized to regions of space in relation to the entire space is formally a map that attaches ``data'' to open sets of a topological space, called a \textit{presheaf} (definition~\ref{defn:presheaf}), which is a kind of functor (remark~\ref{rem:presheaf}). A simple example involves a sequence of visual displays as one might experience in a psychophysics experiment involving stimuli presented on different regions of the display as a presheaf (example~\ref{exmp:presheaf}). The passage from local to global assignment involves unions of open sets to cover wider regions of space (definition~\ref{defn:open cover}) up to the entire space (remark~\ref{rem:open cover}), i.e. the global sections of the presheaf. If just enough global sections are obtained from ``patching'' local data that agree on the intersections of their open sets, then the presheaf is a \textit{sheaf} (definition~\ref{defn:sheaf}), such as the gluing of stimuli as they appear to coincided on an overlapping region of the display (example~\ref{exmp:sheaf}). A sheaf is a universal construction satisfying a certain universal mapping property---unique-existence condition (remarks~\ref{rem:sheaf}). A presheaf that satisfies both uniqueness and existence conditions is a sheaf (remark~\ref{rem:presheaf-condition}). The inclusion/restriction maps are constraints on the unique-existence conditions. Presheaves and sheaves constitute categories (remark~\ref{rem:presheaf-category}).

\subsubsection{Categorical axioms for phenomenal space (revisited, again)}

This local-to-global relationship is illustrated with the adage, ``Red sky at night, sailor's delight. Red sky in morning, sailor take warning.'', in terms of presheaves, depicted as tables (Figure~\ref{fig:adage}). Presheaves can be expressed as relational database tables \cite{abramsky2013relational}, i.e. the table headings correspond to points of the topological space and the table of values for each row to the sections on open sets for the corresponding columns---a row of the table corresponds to a global section. Regarding the adage, then, the headings of the (two-column) table constitute the western and eastern regions of the sky (as points of the topological space) and the rows express the colours (for each region) in the evening and in the morning (as the attached data). The local-to-global relationship is illustrated by the restriction (projection) of the table along its West and East columns yielding the colours of that region for evening and morning. The values on columns are the sections for the corresponding open sets and each row (of the two-column table) is a global section for the whole space. The adage is essentially conveying a potential implication for the sailor, hence the associated feelings of delight and warning (foreboding) as the global sections and their restrictions to feelings for each region as tables for the same space, which is a presheaf morphism. (For more examples of the database perspective on presheaves, see \citeNP{abramsky2011sheaf}, in the context of measurement, and \citeNP{phillips2018going, phillips2019sheaving}, in the context of cognition; see also \citeNP{rosiak2022sheaf}.) 

\begin{figure}[ht]
	\begin{equation*}
		\begin{tabular}{|c|c|c|c|c|c|c|ccccc} 
			\cline{1-1} \cline{3-4} \cline{6-6}
			red & $\mapsfrom$ & red & black & $\mapsto$ & black & \multicolumn{5}{l}{\hspace{5mm}``Red sky at night''} \\ 
			\cline{1-1} \cline{3-4} \cline{6-6}
			black && black & red && red & \multicolumn{5}{l}{\hspace{5mm}``Red sky in morning''} \\ \cline{1-1} \cline{3-4} \cline{6-6}
			
			\textbf{west} & $\rightarrow$ & \textbf{west} & \textbf{east} & $\leftarrow$ & \textbf{east} \\ \cline{1-1} \cline{3-4}  \cline{6-6} 
			\multicolumn{6}{c}{$\Downarrow$} \\
			
			\cline{1-1} \cline{3-4} \cline{6-6}
			\textit{warm} & $\mapsfrom$ & \multicolumn{2}{c|}{\textit{delight}} & $\mapsto$ & \textit{.} & \multicolumn{5}{l}{\hspace{5mm}``sailor's delight''} \\ 
			\cline{1-1} \cline{3-4} \cline{6-6}
			\textit{.} && \multicolumn{2}{c|}{\textit{warning}} && \textit{dire} & \multicolumn{5}{l}{\hspace{5mm}``sailor take warning''} \\ \cline{1-1} \cline{3-4} \cline{6-6}
			
			\textbf{west} & $\rightarrow$ & \textbf{west} & \textbf{east} & $\leftarrow$ & \textbf{east} \\ \cline{1-1} \cline{3-4}  \cline{6-6} 
		\end{tabular} 
	\end{equation*}	
	\caption{The adage ``Red sky at night, sailor's delight. Red sky in morning, sailor take warning.'' as a presheaf, expressed in the form of relational database tables. (``.'' stands for some nondescript feeling.) The arrows $\rightarrow$ and $\leftarrow$ are inclusions of open sets and the arrows $\mapsfrom$ and $\mapsto$ are corresponding restrictions as applied to a global section. (The arrows are reversed in direction by contravariance of presheaves.)}\label{fig:adage}
\end{figure}

In formal terms, the topological space is a two-point space with points corresponding to west and east regions, $\mathit{Sky} = \{W, E\}$, that has the discrete topology, $\{\emptyset, \{W\}, \{E\}, \mathit{Sky}\}$. The data include colour for regions of the sky as black, $B$, and red, $R$. A presheaf for this situation associates colours to regions and the global sections correspond to times of the day expressed as the pair $(R, B)$ for evening, which says that the western sky is red and the eastern sky is black, and the pair $(B, R)$ for morning, which says that the eastern sky is red and the western sky is black. (Each pair restricts to the first and second elements, which are sections on west and east, respectively.) This presheaf for colour can be associated with a phenomenal presheaf whose global section is \textit{delight} and \textit{warning}, respectively, as a presheaf morphism. These global sections may have restrictions to west and east as distinctive aspects of those experiences, \textit{delight} may restrict to \textit{warm} for the west and \textit{warning} to \textit{dire} for the east. The adage as a presheaf pertains to two moments of experience. Each moment is a \textit{subobject} (definition~\ref{defn:subobject}), i.e. the category theoretical analog of subset. A subobject in the context (category) of functors is a \textit{subfunctor} (example~\ref{exmp:subfunctor}), hence in the context of presheaves is a \textit{subpresheaf}, which is a subset of rows in the context of presheaves as tables. So, each moment of experience pertains to a map between subpresheaves, i.e. a subpresheaf morphism. (This categorical analog of subset extends to other kinds of substructures, remark~\ref{rem:subobject}.)

Note that the ``Red sky'' example involves presheaves that are \textit{not} sheaves, because both presheaves lack the (necessary and sufficient) global sections to be sheaves. Intuitively, the presheaves do not say anything about the moments when \textit{both} western and eastern regions of the sky are red, or are black, as neither does the adage. The adage ``excludes'' these two cases. Technically, there does not exist a global section that restricts to red as the (local) section on west \textit{and} red as the section on east---row $(R, R)$, nor a global section that restricts to black (west) \textit{and} black (east)---row $(B, B)$; likewise for the (codomain) presheaf for \textit{delight} and \textit{warning}. Note, also, that the global sections for the first presheaf appear as juxtapositions of the sections to which they restrict, which is a akin to the notion of \textit{classical compositionality}, whereby the constituent (symbols) are ``tokened'' (instantiated) whenever the complex symbol is tokened \cite<see>{fodor1988connectionism}. For the second presheaf, however, the global sections do not appear this way, as \textit{warm} is not so a ``constituent'' of \textit{delight}. For a conceptual distinction, contrast ``black ball'' (i.e. a ball that is black in colour) vs. ``blackball'' (i.e. to ostracize)---the meaning of the latter, by contrast, is not composed from the meanings of ``black'' and ``ball''. The second situation is more closely associated with the IIT notion of ``integration'' \cite{albantakis2023integrated, iit2024wiki, oizumi2014phenomenology}. These differences pertain to presheaves vs. sheaves and natural projections vs. restriction maps more generally, which are taken up in more detail in the supplementary material (Appendix~B).

\subsubsection{Assessment of the axioms}

The move from topological spaces to presheaves---functors \textit{on} topological spaces---reveals a distinction between composition, integration and exclusion as a ``second dimension of variation'' when presheaves are viewed as relational database tables. The table of rows for the whole space (i.e. the global sections) compose ``horizontally'' by inclusions of columns (cf. composition) and amalgamate ``vertically'' as \textit{one} set of rows (cf. integration), but not all possible rows (cf. exclusion). So, phenomenal experience corresponds to a presheaf (existence), which is internal to a presheaf category (intrinsicality). The global sections are specific (information) as one set (integration) and distinguished from other possible presheaves and sections that are not global sections (exclusion). A presheaf may have zero, one, or many global sections, but in every case the collection of global sections for a presheaf is just one set. Each global section is composed of the sections on the corresponding open sets that constitute the underlying topology and the relations that bind these sections are the restriction maps as determined by the action of the presheaf (functor) on the inclusions between corresponding open sets. The discrete and indiscrete spaces are the two extreme topologies for the given set, hence the ``most'' and ``least'' forms of composition, respectively. Regardless of the topology, a presheaf includes a map to just one set of global sections, which may be the empty set, a singleton set, or otherwise. In this sense of there being ``one whole'', the integration axiom pertains to the set of global sections (but see also below) and the composition axiom to how each global section relates to their (local) sections via restriction maps as determined by the topology. In this way, integration and composition axioms are distinguished via locality. Their relationships to UMPs are sketched next.

A presheaf is terminal in its category of subpresheaves given by a \textit{subobject classifier} (definition~\ref{defn:classifier}, example~\ref{exmp:classifier}). The classifier for presheaves says to what extent the data are locally vs. globally ``true'' \cite<see>[for more details regarding an application to cognition.]{phillips2024category} Thus, the set of global sections can be determined as the maximum subobject---cf. the open set $X$ is the terminal subspace in the collection of subspaces for $X$. The subobject classifier and the maximum subobject are both universal constructions, hence integration pertains to UMPs. The presheaf with just one section for each open set is the \textit{terminal sheaf}, i.e. the terminal object in the (sub)category of sheaves on that space, as another kind of universal construction. The collection of subobjects (and their subobject-object relations) to the terminal sheaf is isomorphic to the underlying topological space \cite<see, e.g.,>{leinster2011informal}, so the phenomenal extendedness of space is incorporated with the move to presheaves by recovering the topological space and not any other space, which is a UMP (exclusion). For a given topological space, the category of sheaves is a subcategory of the category of presheaves that are related by another kind of UMP, called an \textit{adjunction} (definition~\ref{defn:adjunction}, examples~\ref{exmp:adjunction}). Adjunction was proposed as a formal relationship between qualia and reports \cite{tsuchiya2023adjunction}. A pair of left and right adjoint functors are universal with respect to each other and ``dual'' in the sense of opposed (remark~\ref{rem:opposing}). The adjoints compose as an \textit{endofunctor}, i.e. a functor to/from the same category (remark~\ref{rem:adjoints}), hence the universality-duality principles in the form of adjunctions constitute maps to/from the system (intrinsicality). So, the five axiomatic properties pertain to UMPs, in primal or dual form; in local or global context, starting with the assertion that experience is a presheaf (existence), thus completing a categorical formalization of the axioms (see Figure~\ref{fig:iit ct}).

IIT says that phenomenal topological structure derives from the intrinsic causal relations among subject units. A central aspect of causal relations is temporal order, i.e. saying ``A causes B'' generally means that $A$ occurs before, or at least not later than $B$, as well as counterfactural information determined from interventions (\citeNP{pearl2009causality}; see also \citeNP{otsuka2024process} for a categorical view). Ordered sets are closely related to topological spaces (example~\ref{exmp:adjointness}). So, a sequence of causal relationships such as ``A causes B'', ``B causes C'' is associated with a topological space by another adjoint situation. From our category theory perspective, the emergence of phenomenal topological structure from intrinsic causal relations corresponds to an adjoint functor, which is a universal construction.

There are some caveats regarding this formalization. Up to this point, little has been said about categorical connections to postulates, as the focus here has been on the axioms, as a way around the potential pitfalls of operational approaches (see Introduction). Nonetheless, the postulates make more specific claims about the axiomatic properties, such as the integration and exclusion axioms, so a closer comparison is given in the supplementary material (Appendix~B).

\section{Discussion}\label{sect:discussion}

The categorical approach to axiomatizing consciousness presented here helps clarify the status of IIT axioms by relating them to meta-theoretical principles: IIT axioms and principles pertain to various UMPs. Note that this connection is not reductionist, as the UMPs are of different kinds. Rather, the general point here is to show how these axioms and principles can be made formally precise for comparison within IIT and across other theories and applications, as discussed next.

\subsection{Some convergences and divergences}

This categorical approach converges on the axioms, but diverges on their relationships. Recall that a central feature of IIT axioms/postulates is their ordering from existence to composition. For the postulates, as a computational procedure for determining the quantity and quality of experience, each step must be taken in order, thereby imposing a (linear/total) ordering on the postulates and therefore the axioms as their operational realization. Axioms for mathematical structures are not ordered this way, beyond for convention or conceptual clarity. The corresponding category theory versions are not ordered. However, (co)limits are closely related by the shape of the diagrams. Moreover, every finite limit is constructed from just two kinds of limits, i.e. products and equalizers (equivalently, terminals and pullbacks). Products and equalizers are limits to diagrams whose shapes are a pair of objects and a pair of arrows (equivalently, terminals and pullbacks are limits to diagrams whose shapes are the empty category and a pair of converging arrows), respectively. Dually, this situation applies to finite colimits. So, the UMPs for (co)limits are related by shape, but not necessarily order.

Another distinctive aspect of IIT is the unfolding of cause-effect structures at any desired spatial or temporal \textit{grain} \cite{iit2024wiki}, i.e. the base level of physical units in the candidate complex considered in regard to, say, the \textit{neural correlates of consciousness} (NCC) at the level of neurons and their spike-trains of activity \cite<see>[for a review of NCC]{koch2016neural}. IIT identifies a physical system with its system of intrinsic cause-effect relations, though this identification need not be a one-to-one correspondence, hence the distinction between micro-level and macro-level interactions \cite{hoel2016can, iit2024wiki}. The maximum and minimum principles of existence employed by IIT may be comparable to limits and colimits, such as greatest lower and least upper bounds in the case of ordered sets, respectively. Limit and colimit functors are, respectively, right and left adjoints to diagonal functors. So, this kind of correspondence between phenomenal and physical states suggests another adjoint situation involving a category for the physical system and a category for the intrinsic causal relations. (A sheaf theory approach has also been taken towards modelling neural systems, \citeNP{barbero2022sheaf, hansen2020sheaf}.) IIT says that experience pertains to relations that take and make a difference to the system itself. This situation compares with the fact that (left and right) adjoints compose as endofunctors.

Some notion of ``level'' appears at various places in the IIT and category theory views. For IIT, intrinsic information (intrinsicality postulate) is computed between units within a complex, whereas integrated information (integration postulate) is computed with respect of partitions (sets) of units. Hence, IIT has a sense of levels as causal relations within vs. between sets of units. The category theory perspective presented here also has notions of levels in the (higher) categories sense of objects as categories in other categories and functors as arrows that map arrows. Phenomenal experience involves intra-subsystem and inter-subsystem relations, as one might expect if composition of experience is taken as axiomatic.

Although there are similarities and differences between IIT and this categorical approach, they are not supposed to be directly comparable/contrastable as competing theories since one (IIT) is supposed to be a theory and the other (category theory) is supposed to be a meta-theory. 

\subsection{Beyond IIT}

The value of a meta-theory is supposed to be in the relations between theories, like the value of category theory with respect to mathematics. In this regard, there are a number of ways that this categorical perspective may go beyond IIT, depending on what one takes to be the base level. For instance, an important implication of IIT pertains to changes in phenomenal experience with changes in the underlying neural substrate, such as the result of a scotoma \cite{olcese2024accelerating}---phenomenal distance should change with changes in the causal relations induced by brain damage \cite{ellia2021consciousness, haun2019does}. This prediction is currently being actively tested empirically in the aforementioned adversarial collaboration, in contrast to predictive processing accounts of consciousness. Note, however, this difference is obtained from specifying two different substrate models. One way that the categorical approach can go beyond IIT is by capturing the relationship between such models. Topological spaces are related by continuous functions, and every continuous function induces a pair of adjoints functors between the corresponding categories of presheaves (or, sheaves). So, the categorical (meta-theoretical) approach says that such changes are related by functors that dilate space. This situation compares with the cognitive psychological notion of chunking/dechunking of concepts in terms of functors between categories of sheaves \cite{phillips2021category}.

A similar consideration appears in the application of IIT to the phenomenal (directional) flow of time \cite{comolatti2024time}. The phenomenology of space \cite{haun2019does} and the phenomenology of time result from applying the same postulates to unfold from different substrate models the different cause-effect structures that are supposed to correspond to the different qualities of space versus time. The model for time is closely related to the model for space but as a kind of ``directed'' topological space \cite{comolatti2024time}. From a category theory perspective, this situation suggests an adjoint situation, whereby certain topological spaces are related to certain ordered sets in one direction and ordered sets to topological spaces in the other direction as a pair of adjoint functors, thus giving a phenomenal direction to time via a UMP (see example~\ref{exmp:adjointness}).

The current approach was inspired by a parallel between the meta-problem of consciousness \cite{chalmers2018meta, chalmers2020can} and category theory as meta-mathematics \cite{eilenberg1945general}. As a meta-theory, then, this categorical approach may shed some light on the meta-problem, which is to explain why the problem of conscious appears to be a ``hard'' problem in contrast to the meta-problem of consciousness, which appears to be ``easy'' \cite{chalmers2018meta, chalmers2020can}. One might simply appeal to a schema-like fact in the sense that the number of schemas is much smaller than the number of instances of those schemas---cf. relational databases where (by design) there is just one meta-schema for storing all possible schemas for any state of a database \cite{nijssen1989conceptual}. However, the asymmetry in the meta-problem is that belief states, e.g., my belief that apples are red, do not appear to engender the same kind of ``introspective opacity'' as phenomenal states, e.g., the experience of a red apple \cite<see>[for rebuttals of theories, including IIT, around this point]{chalmers2018meta, chalmers2020can}. The categorical approach presented in the current work is suggestive as the difference between topologies on finite versus infinite sets (e.g., the interval topology on the real numbers). In the latter case, there are usually no singleton open sets, so one must take a colimit of a sequence of ever smaller neighbourhoods containing the point of interest to determine the sections (data)---cf. the ``unfathomable richness of seeing'' \cite{haun2024unfathomable}. Presheaves on infinite spaces cannot be given by a relational database schema. By contrast, conceptual knowledge as constructed on finite spaces, with data on individual points as the only elements of open sets, may be assessed directly without taking limits---cf. a sheaf theoretic approach to the Language of Thought \cite{phillips2024category}. Indeed, this difference is exploited for practical applications of sheaf theory to sensing \cite{robinson2014topological}.

A meta-theory is supposed to be about a theory of theories and so should also say something about how extant theories connect. In regard to mathematics and computer science where existing theories are well-established the role of category theory is exacting. In regard to consciousness where there is much less agreement on foundations the role of category theory may be more facilitatory or conciliatory where questions arise. For instance, how is IIT with a focus on phenomenology supposed to connect to other theories of consciousness \cite{del2021comparing, doerig2021hard, seth2022theories, signorelli2021explanatory}? What is the connection between phenomenal structure and perceptional/conceptual structure? For a distinction, perceptual states are triggered by external stimuli, whereas phenomenal states may also be elicited by dreams and hallucinations \cite{mayner2024intrinsic}. To illustrate, category/sheaf theory has been used to assess compositional perceptual/conceptual structure \cite{phillips2018going} in a series of cue-target learning tasks \cite{phillips2016failures}. The basic idea is that conceptual structure could be assessed in terms of generalization (correct responses) on novel cues (pairs of alphabetic characters). Above chance performance was observed for participants who reported awareness of the underlying product structure of the task. Changes in awareness over the course of learning corresponds to another kind of universal construction pertaining to the relationship between presheaves and sheaves, called \textit{sheaving} or \textit{sheafification}, that involves changes in the underlying topology by a continuous function \cite{phillips2018going}. 

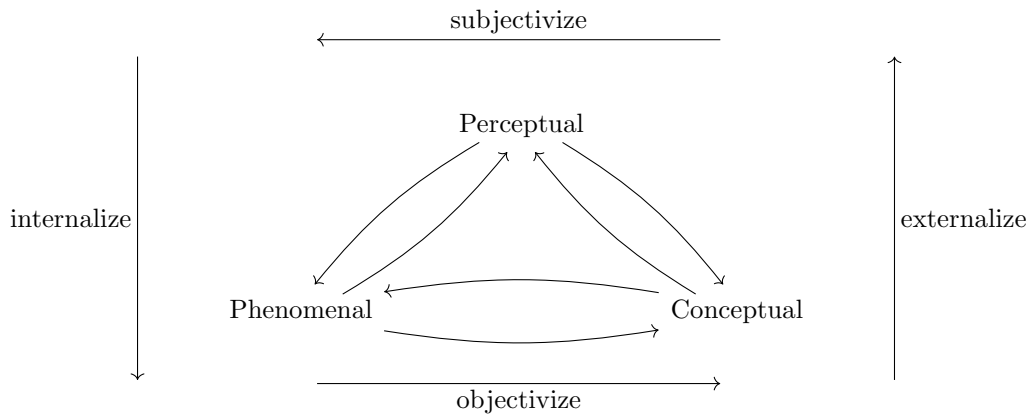
\begin{figure}[ht]
	\begin{equation*}
		\begin{tikzcd}
			\mbox{\ } \arrow[ddddd, "\mbox{internalize}"'] & \mbox{\ } \arrow[rr, "\mbox{subjectivize}", <-] && \mbox{\ } & \mbox{\ } \arrow[ddddd, "\mbox{externalize}", <-] \\
			& & \mbox{Perceptual} \arrow[lddd, shift left, <-, bend left=10] \arrow[lddd, shift right, bend right=10] \arrow[rddd, shift left, bend left=10] \arrow[rddd, shift right, <-, bend right=10]\\ \\ \\
			& \mbox{Phenomenal} \arrow[rr, shift left, <-, bend left=9] \arrow[rr, shift right, bend right=9] && \mbox{Conceptual}  \\
			\mbox{\ } & \mbox{\ } \arrow[rr, "\mbox{objectivize}"'] && \mbox{\ } & \mbox{\ } 
		\end{tikzcd}
	\end{equation*}	
	\caption{Cognitive systems as involving perceptual, conceptual and phenomenal (sub)systems related by adjoint functors (curved arrows) as framed by two ``dimensions'' of opposing drives: subjectivize vs. objectivize and internalize vs externalize.}\label{fig:cognitive system}
\end{figure}

Such transitions suggest ways of using this categorical approach to link phenomenal and conceptual structure and theories thereof---cf. for linking perceptual and conceptual structure in regard to a Language of Thought \cite{phillips2024category}. The phenomenal system is framed in relation to perceptual and conceptual systems as adjoint situations in two ``dimensions'' of oppositions: (1) subjective vs. objective and (2) internal vs. external  (Figure~\ref{fig:cognitive system}). A (re)conciliation of opposing views ``horizontally'' has subjective-first (IIT) on one side and objective-first (the predominant view of science) on the other \cite<see also>[p.~84--85]{lawvere2009conceptual}, and ``vertically'' has psychology as a science of \textit{mental states} (i.e. states that are ``about'' other states) in contraposition to the other sciences absolved of such desiderata \cite{wilson1999philosophy}. Adjoint relations are seen as important in bridging the gap between the phenomenal and the reportable \cite{tsuchiya2023adjunction}.

\subsection{Backing up looking forward}

The approach presented here is primarily about (categorical) universal constructions, vis-a-vis, universal mapping properties, which begs a broader question of why take this approach at all. The very short answer is another apparent parallel with explanations for \textit{systematicity} properties of cognition, i.e. an explanation for why having certain cognitive abilities implies having certain other cognitive abilities. The classical (symbol system) explanation appeals to a particular notion of compositionality \cite{fodor1988connectionism}, but is unclear on why a cognitive system should be compositional in that way \cite{aizawa2003systematicity}. The categorical explanation says that such forms of compositionality follow from the relevant universal constructions \cite{phillips2010categorial}, which are themselves determined by a kind of structural ``gradient descent''---a universal construction is a terminal (dually, initial) object in the relevant category and so is obtained from any object in the category by arrow-traversal as the structural analog of a measure space with a global minimum but no local minima \cite<see>[for perspectives and a formal procedure]{phillips2021category, phillips2016systematicity}. In this way, one may see the existence of consciousness itself (existence axiom) as following from the search for a universal construction, rather than just an assertion, as may arise in a moment of awakening. (This parallel with the systematicity problem, though, does not suppose that conceptual and phenomenal structure are likewise compositional.)

On the other side of this parallel, however, is the problem of establishing (empirically and theoretically) just how ``universal'' is universal. A universal construction in one category may not exist in another. A brute-force approach, akin to the IIT postulates, is to start with a ``large'' category (candidate complex), iteratively try all possible subcategories and compares universal constructions by some measure. In the absence of brute-force, one may encounter ``local minima'' as universal constructions in one category may be less than ideal compared to universal constructions in another category. Alternatively, one may appeal to a notion of \textit{narrow} vs. \textit{broad qualia} \cite{balduzzi2009qualia, kanai2012qualia} as data on open sets (narrow) vs. the total set (broad) and examine whether the narrow qualia can be ``patched'' together as the sheaf condition \cite{youngzie2024towards}, which is a UMP. \cite<See>[for other category theory approaches to IIT.]{northoff2019mathematics, tull2020integrated}

Categorical axiomization of presheaves and sheaves makes clear that one need not be restricted to set-valued (pre)sheaves---a category with products and equalizers (or, pullbacks and terminal objects) suffices, e.g., a category of vector spaces or measure spaces, which may be useful for axiomatizing other kinds of phenomenal experience. The ``atoms'' of phenomenal experience need not be just sets of distinctions that take/make a difference \cite{iit2024wiki}, but rather spaces in their own right. One might hope that this kind of formal axiomatization is extendable towards a ``periodic table'' for organizing qualia structures \cite{tsuchiya2024qualia}. Casting IIT axioms in terms of UMPs is suggestive as finite (co)limits are related by the shapes of their diagrams. In this way, qualia may organize around the shapes of diagrams or topological spaces.

The IIT postulates determine when an internal causal relation constitutes a phenomenal experience and so IIT does not imply \textit{panpsychism} \cite{skrbina2003panpsychism}---``Unlike panpsychism, however, IIT clearly implies that not everything is conscious.'' \cite{tononi2015consciousness}. In like manner, not every mental state is supposed to be a syntactically/semantically compositional state \cite{fodor1975language, fodor1988connectionism}; not every construction in category theory is a universal construction. Despite the broad applicability of category theory and the oft-quoted slogan, ``adjoints are everywhere'' \cite<>[p.~97]{maclane1998categories}, the approach presented here is not supposed to claim that consciousness is everywhere. Rather, universal mapping properties, like the sheaf condition, afford additional tests for assessing such differences. In this spirit, proponents of IIT may regard this approach as a guide towards developing more computationally tractable methods for computing the quantity and quality of consciousness experience as formalized by the postulates. 

On a closing note, the crux of the problem in proffering a scientific theory of consciousness turns on the apparent paradox of having to claim an objective (scientific) theory for something that is essentially subjective, i.e. first-person, phenomenal experience \cite{ellia2021consciousness}. The category theory concept of adjoint situation, as it pertains to universal mapping properties in the form of universal morphisms, appears to offer a general way around this conundrum. Recall, for a pair of functors $F : \mathbf{C} \rightarrow \mathbf{D}$ and $G : \mathbf{D} \rightarrow \mathbf{C}$, the adjoint situation $F \dashv G$ says that there is a universal morphism from object $X$ to $G$ for every $X$ in $\mathbf{C}$ and a universal morphism from $F$ to object $Y$ for every $Y$ in $\mathbf{D}$. Recall, also, that a universal morphism factors every arrow into the composition of a common (mediating) arrow and a unique arrow: for the adjoint situation, the universal morphism from $X$ to $G$ says that every arrow $f : X \rightarrow G(A)$ factors as $f = G(u) \circ \eta_X$, where $G(u)$ and $\eta_X$ are the unique and mediating arrows, respectively (and likewise for the universal morphisms from $F$ to $Y$). The mediating component is regarded as the invariant (``objective'') component and the unique arrows are the variated (subjective) components. If $\mathbf{C}$ is supposed to be a category of subject experiences and $\mathbf{D}$ a category of events, then the universal morphisms from $X$ to $G$ capture the invariant (objective aspect) of the experiences for those events as the corresponding mediating arrow, $\eta_X$. In this way, the subjective at one level (UMP internal to a category) is reconciled with the objective at another level (UMP external to that category), so the importance of a having a category (meta-)theoretical account of consciousness, whence the slogan, ``Consciousness is a universal (mapping) property.''

\paragraph{Acknowledgment}
We thank participants of the Qualia Structure Summer School for discussions, especially Giulio Tononi, Francesco Ellia, Matteo Grasso, Michael Robinson and Hayato Saigo. An earlier version of this work was presented at the ASSC27 meeting \cite{phillips2024towards}. This work was supported by Japanese Society for the Promotion of Science Grant-in-aid (23H04829, 23H04830). NT was also supported by the National Health and Medical Research Council (APP1183280) and the Australian Research Council (DP240102680). This research was also conducted while NT was visiting the Okinawa Institute of Science and Technology (OIST) through the Theoretical Sciences Visiting Program (TSVP). 

\paragraph{Data availability statement}
There are no new data associated with this article.


\newpage
\appendix

\section{Basic theory}

\subsection{Sets}

\begin{defn}[topological space]\label{defn:topological space}
	A \textit{topological space} is a pair $(X, T)$ consisting of a set $X$ and set $T$ of subsets of $X$, called the \textit{topology} of $X$, whose elements $U \in T$ are called the \textit{open sets} of $T$, such that
	\begin{itemize}
		\item the empty set and $X$ are open sets, i.e. $\emptyset, X \in T$,
		\item finite intersections of opens sets are open sets, i.e. $(U_i \in T)_{i \in I} \Rightarrow (\bigcap_i U_i) \in T$, $I \subseteq \mathbb{N}$, and
		\item arbitrary unions of opens sets are open sets, i.e. $(U_i \in T)_{i \in I} \Rightarrow (\bigcup_i U_i) \in T$, $I \subseteq \mathbb{N}^\infty$.
	\end{itemize}
	Set $X$ is also called the \textit{total set}.
\end{defn}

\begin{rem}\label{rem:topological space}
	Inclusion relations capture a notion of proximity, e.g. $\{x, y\} \subseteq \{x, y, z\}$ says that points $x$ and $y$ are closer to each other than to $z$. Every set $X$ has two (extreme) topologies:
	\begin{itemize}
		\item \textit{discrete}: the power set---every subset is an open set, and
		\item \textit{indiscrete}: only the empty set and $X$ are open sets.
	\end{itemize}
	The discrete space conveys the sense that every element in the space is equally proximal; the indiscrete space conveys no information about proximity.
\end{rem}

\begin{defn}[partially ordered set]\label{defn:poset}
	A \textit{partially ordered set}, also called a \textit{poset}, is a pair $(P, \le)$ consisting of a set $P$ and a relation $\le$ on $P$, called a \textit{partial order}, that is
	\begin{itemize}
		\item \textit{reflexive}: $p \le p$ for all $p \in P$,
		\item \textit{antisymmetric}: $p \le q$ and $q \le p$ implies $p = q$ for all $p, q \in P$ and
		\item \textit{transitive}: $p \le q$ and $q \le r$ implies $p \le r$.
	\end{itemize}
	
\end{defn}

\begin{rem}\label{rem:poset}
	A topological space is a poset, as inclusion is reflexive ($U \subseteq U$), antisymmetric ($V \subseteq U$ and $U \subseteq V$ implies $V = U$) and transitive ($W \subseteq V$ and $V \subseteq U$ implies $W \subseteq V$).
\end{rem}

\begin{defn}[metric space]\label{defn:metric space}
	A \textit{metric space} is a pair $(M, d)$ consisting of a set $M$ and a \textit{distance} function $d : M \times M \rightarrow \mathbb{R}$ that is
	\begin{itemize}
		\item indiscernible: $d(x, x) = 0$ for all $x \in M$,
		\item symmetric: $d(x, y) = d(y, x)$ for all $x, y \in M$ and 
		\item subadditive (triangle inequality): $d(x, z) \le d(x, y) + d(y, z)$ for all $x, y, z \in M$.
	\end{itemize}
	
\end{defn}

\begin{rem}\label{rem:metric space}
	Every metric space has a topology constructed from \textit{open balls}.
\end{rem}

\begin{defn}[measurable space]\label{defn:measurable space}
	A \textit{measurable space} is a pair $(X, \Sigma)$ consisting of a set $X$ and a $\sigma$-algebra $\Sigma$, i.e. a set of subsets of $X$ such that
	\begin{itemize}
		\item $X$ is in $\Sigma$ and is closed under
		\item complementation: if $A$ is in $\Sigma$, then $X \setminus A$ is in $\Sigma$ and
		\item countable unions: $\bigcup_i^\infty A_i \in \Sigma$.
	\end{itemize}
\end{defn}

\begin{rem}\label{rem:measurable space}
	The underlying set of a finite $\sigma$-algebra corresponds to a topology where the elements of the underlying set of $\Sigma$ are the open sets. However, a complement of an open set need not be an open set for a topological space. For comparison, $\Sigma$ consists of:
	\begin{itemize}
		\item the empty set and $X$,
		\item countable intersections,
		\item countable unions and 
		\item complementary sets.
	\end{itemize}
	The axioms for a measurable space imply that the empty set and countable intersections are in $\Sigma$.
\end{rem}

\begin{defn}[measure space]\label{defn:measure space}
	A \textit{measure space} is a measurable space equipped with a \textit{measure}, i.e. a function $\mu : \Sigma \rightarrow \mathbb{R}$ such that
	\begin{itemize}
		\item $\mu(\cup s_i) = \Sigma_i \mu(s_i)$, for all countable pairwise disjoint unions,
	\end{itemize}
	together written as the triple $(X, \Sigma, \mu)$.
\end{defn}

\begin{rem}\label{rem:probability space}
	If $\mu(X) = 1$, then the measure space is a \textit{probability space}.
\end{rem}

\subsection{Categories}

Category theory has two views on ``structure'' relative to a category: internal (section~\ref{sect:internal}) and external (section~\ref{sect:external}). Constructions defined or determined by universal mapping properties are universal constructions and so too their dual constructions (section~\ref{sect:universal}). The distinction between internal vs. external is at least tacit in introductory texts on category theory, e.g., ``Limits are about what goes on \textit{inside} a category.'' \cite<>[p.~107, emphasis there]{leinster2014basic}, where the concept of \textit{limit} is subsequently revisited from the perspective of \textit{adjoint functors} \cite<>[ch.~6]{leinster2014basic}. Such instances of providing definitions for a formal concept from multiple perspectives are commonplace in category theory \cite<see also, e.g.,>[]{maclane1998categories}. The emphasis, here, on this distinction is in light of the intrinsicality axiom of IIT \cite{albantakis2023integrated, iit2024wiki}, where the difference between internal vs. external causal relations plays a crucial role in determining the quantity and quality of experience. Other introductions to the category theory, sheaf theory and topos theory concepts used here are also available \cite<see, e.g.,>[]{goldblatt2006topoi, lawvere2009conceptual, maclane1992sheaves, rosiak2022sheaf}. For an introduction to higher category theory see here \cite{grandis2020higher}.

\subsubsection{Internal structure to a category}\label{sect:internal}

\begin{defn}[category]\label{defn:category}
	A \textit{category} is a tuple $\mathbf{C} = (\mathbf{C}_0, \mathbf{C}_1, \id, \dom, \cod, \circ)$ consisting of:
	\begin{itemize}
		\item a collection of entities, called \textit{objects}, $\mathbf{C}_0 = \{A, B, C, \dots, \}$,
		\item a collection of directed ``relations'' between objects, called \textit{arrows} (\textit{maps} or \textit{morphisms}), $\mathbf{C}_1 = \{f, g, h, \dots \}$---an arrow written in full as $f : A \rightarrow B$ indicates that $f$ is directed from $A$ to $B$, called the \textit{domain} and \textit{codomain} of $f$, respectively,
		\item a map assigning to each object $A$ an arrow, written $1_A : A \rightarrow A$, called an \textit{identity arrow}, i.e. $\id : \mathbf{C}_0 \rightarrow \mathbf{C}_1; A \mapsto 1_A$
		\item two maps assigning to each arrow $f$ a domain and codomain object:
		\begin{itemize}
			\item $\dom : \mathbf{C}_1 \rightarrow \mathbf{C}_0; f \mapsto A$,
			\item $\cod : \mathbf{C}_1 \rightarrow \mathbf{C}_0; f \mapsto B$ and
		\end{itemize}
		\item a \textit{composition operation}, $\circ$, sending each pair of \textit{compatible} arrows $f : A \rightarrow B$ and $g : B \rightarrow C$ (i.e. the codomain of $f$ is the domain of $g$) to the arrow $g \circ f : A \rightarrow C$ that satisfies
		\begin{itemize}
			\item \textit{unity}: $f \circ 1_A = f = 1_B \circ f$ for every arrow $f$ in $\mathbf{C}$, and
			\item \textit{associativity}: $h \circ (g \circ f) = (h \circ g) \circ f$ for every triple of arrows $f, g, h$ in $\mathbf{C}$.
		\end{itemize}
		
	\end{itemize}
	The collection of arrows in $\mathbf{C}$ with domain $A$ and codomain $B$ is called a \textit{hom-set}, written $\Hom_\mathbf{C}(A, B)$, $\mathbf{C}(A, B)$, or $[A, B]$ when the category is understood.
\end{defn}

\begin{rem}\label{rem:graph}
	A category is similar in structure to a directed graph, i.e. informally, a category consists of a collection of entities, called \textit{objects} (cf. vertices), a collection of directed ``relationships'' between objects, called \textit{arrows}, \textit{maps}, or \textit{morphisms} (cf. edges), and an operation that composes certain arrows into arrows, called \textit{composition} (cf. concatenation of edges to form paths). A formal comparison follows.
	\begin{itemize}
		\item A \textit{directed graph}, $G$, consists of a set of points, $G_0$, called \textit{vertices}, and a set of directed links between vertices, $G_1$, called \textit{edges}---each edge, $e$, goes from one vertex, $s$, called the \textit{source} of $e$, to one vertex, $t$, called the \textit{target} of $e$, written $e : s \rightarrow t$. An edge from/to the same vertex is called a \textit{loop}. A pair of edges, $(e, e')$ is called connected when the target of $e$ is the source of $e'$. A sequence of connected edges is called a \textit{path}. The relationships between vertices and edges is given by two maps that assign to each edge its source and target, i.e. $\mathit{src}, \mathit{tgt} : G_1 \rightarrow G_0$, respectively. Thus, a graph is a tuple $G = (G_0, G_1, \mathit{src}, \mathit{tgt})$.
	\end{itemize}
	Compare object to vertex, arrow (from a domain to a codomain) to edge (from a source to a target), identity arrow to loop, $\dom$/$\cod$ relation to $\mathit{scr}$/$\mathit{tgt}$ relation, and composition of (compatible) arrows to concatenation of (connected) edges. There are, however, two essential differences between categories and graphs. Firstly, every object in a category is associated with an identity arrow, but a directed graph need not have a loop for each vertex. Secondly, for every pair of compatible arrows, $(f, g)$, there is an arrow, $g \circ f$, given by the composition operation, but for every path from a vertex $v$ to a vertex $w$ there need not be a corresponding edge from $v$ to $w$, i.e. there are no maps corresponding to $\mathit{id}$ and $\circ$ in the signature (tuple) for a graph. Any arrangement of entities and their relationships that satisfies those conditions is a category. 
\end{rem}

\begin{exmp}[sets and functions]\label{exmp:sets}
	The category $\mathbf{Set}$ has sets for objects and (total) functions between sets for arrows. Composition is function composition; the identity arrows are the identity functions.
\end{exmp}

\begin{exmp}[sets and inclusions]\label{exmp:inclusions}
	The category $\mathbf{Set}^\subseteq$ has sets for objects and inclusions for arrows. Composition is given by transitivity, i.e. $W \subseteq V$ and $V \subseteq U$ implies $W \subseteq U$, and identities by reflexivity, i.e. $U \subseteq U$. An inclusion $V \subseteq U$ is a map $\iota : V \rightarrow U$ sending every element in $V$ to the same element in $U$, i.e. $\iota : v \mapsto v$. If $V$ and $U$ are the same set, then the inclusion is also an identity, $1_U$.
\end{exmp}

\begin{rem}\label{rem:subcategory}
	$\mathbf{Set}^\subseteq$ is a \textit{subcategory} of $\mathbf{Set}$, the categorical analog of subset: a category $\mathbf{C}$ is a subcategory of a category $\mathbf{D}$ if every object/arrow in $\mathbf{C}$ is an object/arrow in $\mathbf{D}$ and composition is given by the composition operation of $\mathbf{D}$, hence the reuse of notation, $\mathbf{C} \subseteq \mathbf{D}$.
\end{rem}

\begin{exmp}[topological space]\label{exmp:topological space}
	A topological space $(X, T)$ is a category whose objects are the open sets $U \in T$ and arrows are the inclusions $V \subseteq U$.
\end{exmp}

\begin{exmp}[poset]\label{exmp:poset}
	A poset, $(P, \le)$, is a category whose objects are the elements $p \in P$ and arrows are the order relationships, i.e. there is an arrow $p \rightarrow q$ whenever $p \le q$. 
\end{exmp}

\begin{rems}\label{rem:ump}
	Universal constructions come in a variety of forms. Each such construction is determined by a certain \textit{universal mapping property} that is defined in terms of a \textit{unique-existence condition} relating all constructions in the domain of interest to the universal construction. For a basic intuition, the maximum element in an ordered set is a universal construction in that every element in that set is less than or equal to the maximum element. In terms of arrows (hence, mapping properties), for every element there is exactly one arrow from it to the maximum element, so satisfying a unique-existence condition. 
	\begin{itemize}
		\item Each form of universal construction is defined in a case-by-case manner. However, each definition typically follows a familiar pattern, e.g., ``in a category ... a ... is a ... such that for every ... there \textit{exists} a \textit{unique} arrow ... such that ... is satisfied.''
		
		\item For comparison, in a poset $(P, \le)$ a maximum element is an element $\top$ (also called the \textit{top element}) such that for every element $p$ there exists a unique order relationship $p \le \top$, i.e. in other words, every element $p$ in $P$ is less than or equal to $\top$. For ordered sets as categories, uniqueness is automatic since there can be at most only one arrow from an element $p$ to an element $q$ in $P$.
		
		\item The unique-existence condition is often expressed by a so-called \textit{commutative diagram}, whereby every pair of paths of compatible arrows that start at the same object and end at the same object are equal (i.e. the two paths express the same composite arrow), where at least one path consists of at least two arrows (at least one of which is not an identity arrow). Uniqueness is usually expressed by a dashed arrow, as shown in the cases defined below.
		
		\item A universal construction of a particular kind need not exist. For instance, the set of natural numbers with the usual ordering does not have a maximum number.
		
		\item The notion of ``construction'' is in a formal sense, not in an operational sense of being the result of some (computational) procedure. In computational category theory, however, some universal constructions are the basis for specific computational procedures \cite<see, e.g.,>[]{hinze2016unifying}.
	\end{itemize}
	Specific cases of universal constructions that are central to this categorical approach to consciousness are subsequently defined.
\end{rems}

\begin{defn}[product]\label{defn:product}
	In a category $\mathbf{C}$, the \textit{product} of objects $A$ and $B$ (if it exists) is an object $P$, also denoted $A \times B$, together with a pair of arrows $\pi = (\pi_A, \pi_B)$ such that for every object $Z$ and every pair of arrows $z_A : Z \rightarrow A$ and $z_B : Z \rightarrow B$ there exists a unique arrow $u : Z \rightarrow P$ such that $z = \pi \circ u$, as shown by the following commutative diagram
	\begin{equation}\label{eqn:product}
		\begin{tikzcd}
			& Z \arrow[ld, "z_A"'] \arrow[rd, "z_B"] \arrow[d, "u"', dashed]  \\
			A & A \times B \arrow[l, "\pi_{A}"] \arrow[r, "\pi_{B}"'] & B 
		\end{tikzcd}
	\end{equation}
\end{defn}

\begin{exmp}[Cartesian product]\label{exmp:cartesian product}
	In $\mathbf{Set}$, the product is \textit{Cartesian product} of two sets, $A$ and $B$, i.e. the set of all pairs of elements from each set, $A \times B = \{(a, b) | a \in A, b \in B\}$, and two projection maps, $\pi_A : (a, b) \mapsto a$ and $\pi_B : (a, b) \mapsto b$. The unique arrow $u$ is the product arrow, denoted $\langle z_A, z_B \rangle$, which sends each element in $Z$ to the pair of images under $z_A$ and $z_B$, i.e. $\langle z_A, z_B \rangle : z \mapsto (z_A(z), z_B(z))$. 
\end{exmp}

\begin{exmp}[intersection]\label{exmp:intersection}
	In $\mathbf{Set}^\subseteq$, the product is intersection and two inclusions:
	\begin{equation}\label{eqn:intersection}
		\begin{tikzcd}
			A & A \cap B \arrow[l, hook'] \arrow[r, hook] & B 
		\end{tikzcd}
	\end{equation}
\end{exmp}

\begin{defn}[coproduct]\label{defn:coproduct}
	In a category $\mathbf{C}$, the \textit{coproduct} of objects $A$ and $B$ (if it exists) is an object $Q$, also denoted $A + B$, together with a pair of arrows $\iota = (\iota_A, \iota_B)$ such that for every object $Z$ and every pair of arrows $z_A : A \rightarrow Z$ and $z_B : B \rightarrow Z$ there exists a unique arrow $u : Q \rightarrow Z$ such that $z = u \circ \iota$, as shown by the following commutative diagram
	\begin{equation}\label{eqn:coproduct}
		\begin{tikzcd}			
			A \arrow[r, "\iota_A"] \arrow[rd, "z_A"'] & A + B \arrow[d, "u"', dashed] & B \arrow[l, "\iota_B"'] \arrow[ld, "z_B"] \\
			& Z
		\end{tikzcd}
	\end{equation}
\end{defn}

\begin{exmp}[disjoint union]\label{exmp:disjoint union}
	In $\mathbf{Set}$, the coproduct is \textit{disjoint union} of sets $A$ and $B$, i.e. the set of all elements from both sets distinguished by their origin, $A + B = \{(1, a) | a \in A\} \cup \{(2, b) | b \in B\}$, and two injection maps, $\iota_A : a \mapsto (1, a)$ and $\iota_B : b \mapsto (2, b)$. The unique arrow $u$ is the coproduct arrow, denoted $[z_A, z_B]$, which acts as a conditional, i.e.
	\begin{displaymath}
		[z_A, z_B] : (i, x) \mapsto
		\begin{cases} 
			z_A(x)		& \text{if } i = 1 \\ 
			z_B(x)		& \text{if } i = 2
		\end{cases}	
	\end{displaymath}
\end{exmp}

\begin{exmp}[set union]\label{exmp:set union}
	In $\mathbf{Set}^\subseteq$, the coproduct is set union and two injections:
	\begin{equation}\label{eqn:set union}
		\begin{tikzcd}
			A \arrow[r, hook] & A \cup B & B \arrow[l, hook']
		\end{tikzcd}
	\end{equation}
\end{exmp}

\begin{defn}[terminal object]\label{defn:terminal}
	In a category $\mathbf{C}$, the \textit{terminal object} (if it exists) is an object, denoted $1$, such that for every object $Z$ there exists a unique arrow from $Z$ to $1$, written $! : Z \rightarrow 1$.
\end{defn}

\begin{exmp}[singleton set]\label{exmp:singleton set}
	In $\mathbf{Set}$, the terminal object is any singleton set, typically written $\{*\}$ when the identity of the only element is unimportant.
\end{exmp}

\begin{exmp}[total set]\label{exmp:total set}
	The terminal object in a topological space $(X, T)$ is the total set $X$.
\end{exmp}

\begin{defn}[initial object]\label{defn:initial}
	In a category $\mathbf{C}$, the \textit{initial object} (if it exists) is an object, denoted $0$, such that for every object $Z$ there exists a unique arrow from $0$ to $Z$, written $\invexcl : 0 \rightarrow Z$. 
\end{defn}

\begin{exmp}[empty set]\label{exmp:empty set}
	The initial object in $\mathbf{Set}$ and in any topological space is the empty set.
\end{exmp}

\begin{rem}\label{rem:uniqueness}
	Terminals (and other universal constructions) are \textit{unique up to a unique isomorphism}, hence referred to as ``the'' rather than ``a'' universal construction, when this equivalence is all that matters. (An arrow $f : A \rightarrow B$ is an \textit{isomorphism} if there exists an arrow $g : B \rightarrow A$ such that $g \circ f = 1_A$ and $f \circ g = 1_B$.)
\end{rem}

\begin{exmp}[terminal]\label{exmp:equivalence}
	In $\mathbf{Set}$, any singleton set is a terminal object.
\end{exmp}

\begin{rems}\label{rem:duality}
	Dual constructions pertain to reversing the directions of arrows.
	\begin{enumerate}
		\item The category \textit{opposite} to a category $\mathbf{C}$, denoted $\mathbf{C}^\op$, has the objects of $\mathbf{C}$ and the arrows of $\mathbf{C}$ ``reversed'', i.e. an arrow $f : C \rightarrow D$ in $\mathbf{C}$ is the arrow $f^\op : B \rightarrow A$ in $\mathbf{C}^\op$. The arguments to the composition operation are swapped.
		\item A construction in $\mathbf{C}$ is the dual construction in $\mathbf{C}^\op$, e.g., product in $\mathbf{C}^\op$ is coproduct in $\mathbf{C}$. 
		\item Dual constructions are often distinguished from their original (primal) constructions by the ``co'' prefix. 
	\end{enumerate}
	
\end{rems}

\begin{rem}\label{rem:structure categories}
	Structures, e.g., topological spaces, generally form categories in their own right even when that structure is not itself a category, e.g., the collection of (directed) graphs and graph homomorphisms, $\mathbf{Gph}$ is a category. $\mathbf{Top}$ is the category of topological spaces (objects) and \textit{continuous functions} (arrows), i.e. a function $f : X \rightarrow Y$ such that every preimage of an open set in $Y$ is an open set in $X$.
\end{rem}

\subsubsection{External structure to a category}\label{sect:external}

\begin{defn}[functor]\label{defn:functor}
	A \textit{functor} is a map $F : \mathbf{C} \rightarrow \mathbf{D}$ sending each object $C$ in $\mathbf{C}$ to the object $F(C)$ in $\mathbf{D}$ and each arrow $f : C \rightarrow C'$ to the arrow $F(f) : F(C) \rightarrow F(C')$ in $\mathbf{D}$ that satisfies:
	\begin{itemize}
		\item \textit{unity}: $F(1_C) = 1_{F(C)}$ for every object $C$ in $\mathbf{C}$ and
		
		\item \textit{compositionality}: $F(f' \circ_\mathbf{C} f) = F(f') \circ_\mathbf{D} F(f)$ for every pair of arrows $f$ and $f'$ in $\mathbf{C}$.
	\end{itemize}
	$F$ is said to preserve identities and compositions. The collection of objects $F(C)$ and arrows $F(f)$ is called the \textit{image} of $F$ (cf. image of a function).
\end{defn}

\begin{rem}\label{rem:functor}
	A functor is a category homomorphism, cf. graph homomorphism is a pair of maps, $h = (h_0, h_1) : G \rightrightarrows G'$, preserving sources and targets---the source of a mapped edge is the same vertex as the mapped source of the edge, i.e. $\mathit{src'}(h_1(e)) = h_0(\mathit{src}(e))$; likewise, targets.
	
\end{rem}

\begin{exmp}[diagram]\label{exmp:diagram}
	A \textit{diagram} is a functor from a \textit{shape} category $J$ to a category $\mathbf{C}$, i.e. $D : J \rightarrow \mathbf{C}$, that acts like an \textit{indexed set} by picking out a (sub)collection of objects and arrows in $\mathbf{C}$ with shape $J$. 
	\begin{enumerate}
		\item The pair-shaped diagram $(A, B) : 2 \rightarrow \mathbf{C}$ picks out the objects $A$ and $B$ in $\mathbf{C}$. 
		
		\item The vee-shaped diagram $(f, g) : \vee \rightarrow \mathbf{C}$ picks out a pair of arrows $f : A \rightarrow C$ and $g : B \rightarrow C$ in $\mathbf{C}$, i.e. two arrows with the same codomain $C$. 
		
		\item The par-shaped diagram $(f, g) : (\downdownarrows) \rightarrow \mathbf{C}$ picks out a pair of arrows $f, g : A \rightarrow B$ in $\mathbf{C}$. 
		
		\item The empty shape diagram is the empty functor $\emptyset : 0 \rightarrow \mathbf{C}$, i.e. from the category with no objects, hence no arrows (cf. empty function).
	\end{enumerate}

\end{exmp}

\begin{rem}\label{rem:diagram}
	The objects and arrows in the image of $D$ are also denoted $D(i)$ and $D(ij)$, or $D_i$ and $D_{ij}$, respectively (cf. indexed set).
\end{rem}

\begin{defn}[natural transformation]\label{defn:natural transformation}
	Suppose functors $F, G : \mathbf{C} \rightarrow \mathbf{D}$. A \textit{natural transformation} from $F$ to $G$, written $\eta : F \dotrightarrow G$, is a family of \textbf{D}-arrows indexed by the objects $C$ in $\mathbf{C}$, i.e. $\eta = \{\eta_C : F(C) \rightarrow G(C)\}_{C \in \mathbf{C}_0}$, such that for each arrow $f : C \rightarrow C'$ in $\mathbf{C}$ the following diagram commutes:
	\begin{equation}\label{eqn:naturality}
		\begin{tikzcd}
			C \arrow[d, "f"'] & F(C) \arrow[r, "\eta_C"] \arrow[d, "F(f)"'] & G(C) \arrow[d, "G(f)"] \\
			C' & F(C') \arrow[r, "\eta_{C'}"'] & G(C')
		\end{tikzcd}
	\end{equation}
	i.e. $G(f) \circ \eta_C = \eta_{C'} \circ F(f)$. Each arrow $\eta_C$ is called the \textit{component} of $\eta$ at $C$.
\end{defn}

\begin{exmps}[natural transformations]\label{exmp:natural transformations}	
	\leavevmode
	\begin{enumerate}
		\item $A \cap B \subseteq A$ is natural in $A$, as indicated by commutative diagram
		\begin{equation}\label{eqn:naturality intersection}
			\begin{tikzcd}
				A \cap B \arrow[r, "\subseteq"] \arrow[d, "\subseteq"'] & A \arrow[d, "\subseteq"] \\
				A' \cap B' \arrow[r, "\subseteq"'] & A' 
			\end{tikzcd}
		\end{equation}
		More generally, projection is a natural transformation, $\acute{\pi} : A \times B \dotrightarrow A$.
		
		\item $A \subseteq A \cup B$ is also natural in $A$, as indicated by commutative diagram
		\begin{equation}\label{eqn:naturality union}
			\begin{tikzcd}
				A \arrow[r, "\subseteq"] \arrow[d, "\subseteq"'] & A \cup B \arrow[d, "\subseteq"] \\
				A' \arrow[r, "\subseteq"'] & A' \cup B'
			\end{tikzcd}
		\end{equation}
		More generally, insertion is a natural transformation, $\acute{\iota} : A \dotrightarrow A + B$.
		
		\item See \citeA{tsuchiya2021relational} where the left and right visual fields correspond to functors related by a natural transformation.
		
	\end{enumerate}
	
\end{exmps}

\begin{rem}\label{rem:functor category}
	The functors from a category $\mathbf{C}$ to a category $\mathbf{D}$ and their natural transformations constitute a \textit{functor category}, denoted $\mathbf{D}^\mathbf{C}$. For instance, a diagram $D : J \rightarrow \mathbf{C}$ is an object in the functor category $\mathbf{C}^J$. $\mathbf{C}^0 \cong 1$ is the one-object category consisting of the empty functor (cf. singleton set).
\end{rem}

\subsubsection{Universal construction}\label{sect:universal}

\begin{defn}[cone]\label{defn:cone}
	A \textit{cone} to a \textit{J}-shaped diagram $D$ in a category $\mathbf{C}$ is a pair $(V, \phi)$ consisting of an object $V$ in $\mathbf{C}$ and a family of arrows $\phi = \{\phi_i : V \rightarrow D_i\}_{i \in J_0}$ such that the following diagram commutes:
	\begin{equation}
		\begin{tikzcd}\label{eqn:cone}
			V \arrow[r, "\phi_i"] \arrow[rd, "\phi_j"'] & D(i) \arrow[d, "D(ij)"] \\
			& D(j)
		\end{tikzcd}
	\end{equation}
	The object $V$ and the image of $D$ are called the \textit{vertex} and \textit{base} of the cone, respectively. The family of arrows $\phi$ and the objects $D(i)$ are respectively called the \textit{legs} and \textit{feet} of the cone.
\end{defn}

\begin{rem}\label{rem:cone}
	The constant functor picking out the vertex $V$ is identified with the object $V$ in $\mathbf{C}$, hence a cone is given by object $V$, as the domain of functor $V$ does not matter.
\end{rem}

\begin{exmp}[inclusion of intersection]\label{exmp:inclusion of intersection}
	Intersection constitutes the vertex of a cone, whose legs are the inclusions (compare diagram~\ref{eqn:intersection} and diagram~\ref{eqn:cone}).
\end{exmp}

\begin{defn}[cone homomorphism]\label{defn:cone homomorphism}
	Suppose $(V, \phi)$ and $(W, \psi)$ are cones to a \textit{J}-shaped diagram $D$ in a category $\mathbf{C}$. A \textit{cone homomorphism} is an arrow $h : V \rightarrow W$ in $\mathbf{C}$ such that the following diagram commutes:
	\begin{equation}\label{eqn:cone homomorphism}
		\begin{tikzcd}
			V \arrow[r, "\phi_i"] \arrow[rd, "\phi_j", pos=0.3] \arrow[d, "h"'] & D(i) \arrow[d, "D(ij)"] \\
			W \arrow[r, "\psi_j"'] \arrow[ru, "\psi_i"', pos=0.3] & D(j)
		\end{tikzcd}
	\end{equation}
\end{defn}

\begin{exmp}[intersection subset]\label{exmp:intersection subset}
	A subset of an intersection, $Z \subseteq A \cap B$, is a cone homomorphsm---in diagram~\ref{eqn:cone homomorphism} (replace $V$ and $W$ with $Z$ and $A \cap B$, respectively).
\end{exmp}

\begin{defn}[limit]\label{defn:limit}
	Suppose a diagram $D : J \rightarrow \mathbf{C}$. The \textit{limit} to $D$ (if it exists) is a cone $(L, \lim)$ such that for every cone $(V, \phi)$ to $D$ there exists a unique cone homomorphism $u : V \rightarrow L$ such that the following diagram commutes:
	\begin{equation}\label{eqn:limit cone}
		\begin{tikzcd}
			V \arrow[rd, "\phi"] \arrow[d, "u"', dashed] \\
			L \arrow[r, "\lim"'] & D
		\end{tikzcd}
	\end{equation}
\end{defn}

\begin{exmp}[intersection limit]\label{exmp:intersection limit}
	The limit of sets $A$ and $B$ in $\mathbf{Set}^\subseteq$, i.e. the limit to pair $(A, B)$, is their intersection (replace $L$ with $A \cap B$ in diagram~\ref{eqn:limit cone}).
\end{exmp}

\begin{rem}\label{rem:terminal}
	The terminal object is the limit to an empty diagram.
\end{rem}

\begin{rem}\label{rem:limit}
	A collection of cones to a diagram and their cone homomorphisms constitute a category. A limit is a terminal object in a category of cones, i.e. a universal cone. 
\end{rem}

\begin{defn}[universal morphism]\label{defn:universal morphism}
	There are two forms of \textit{universal morphism}:
	\begin{itemize}
		\item A universal morphism \textit{from} an object $X$ in a category $\mathbf{C}$ to a functor $F : \mathbf{D} \rightarrow \mathbf{C}$ (if it exists) is a pair $(A, \alpha)$ consisting of an object $A$ in $\mathbf{D}$ and an arrow $\alpha : X \rightarrow F(A)$ in $\mathbf{C}$ such that for every object $Y$ in $\mathbf{D}$ and arrow $f : X \rightarrow F(Y)$ in $\mathbf{C}$ there exists a unique arrow $u : A \rightarrow Y$ in $\mathbf{D}$ such that $f = F(u) \circ \alpha$, as indicated by the following commutative diagram:
		\begin{equation}\label{eqn:primal universal}
			\begin{tikzcd}
				X \arrow[r, "\alpha"] \arrow[rd, "f"'] & F(A) \arrow[d, "F(u)", dashed] & A \arrow[d, "u", dashed] \\
				& F(Y) & Y
			\end{tikzcd}
		\end{equation}
		
		\item A universal morphism \textit{to} an object $X$ in a category $\mathbf{D}$ from a functor $F : \mathbf{C} \rightarrow \mathbf{D}$ (if it exists) is a pair $(A, \alpha)$ consisting on an object $A$ in $\mathbf{C}$ and an arrow $\alpha : F(A) \rightarrow X$ in $\mathbf{D}$ such that for every object $Y$ in $\mathbf{C}$ and arrow $f : F(Y) \rightarrow X$ in $\mathbf{D}$ there exists a unique arrow $u : Y \rightarrow A$ in $\mathbf{C}$ such that $f = \alpha \circ F(u)$, as indicated by the following commutative diagram:
		\begin{equation}\label{eqn:dual universal}
			\begin{tikzcd}
				Y \arrow[d, "u"', dashed] & F(Y) \arrow[d, "F(u)"', dashed] \arrow[rd, "f"] \\
				A & F(A) \arrow[r, "\alpha"'] & X
			\end{tikzcd}
		\end{equation}
	\end{itemize}
	The second form is dual to the first (primal) form; equivalently, the first form is dual to the second.
\end{defn}

\begin{rem}\label{rem:limit as univeral morphism}
	The \textit{diagonal functor}, $\Delta : \mathbf{C} \rightarrow  \mathbf{C}^J$, sends each object to the constant (\textit{J}-shape) diagram, i.e. $\Delta : C \mapsto (C : J \rightarrow \mathbf{C})$. A limit to a diagram $D : J \rightarrow \mathbf{C}$ is a universal morphism from the diagonal functor $\Delta : \mathbf{C} \rightarrow \mathbf{C}^J$ to $D$, i.e. the pair $(L, \epsilon)$ in the commutative diagram
	\begin{equation}\label{eqn:limit dual universal}
		\begin{tikzcd}
			Z \arrow[d, "u"', dashed] & \Delta(Z) \arrow[d, "\Delta(u)"', dashed] \arrow[rd, "f"] \\
			L & \Delta(L) \arrow[r, "\epsilon"'] & D
		\end{tikzcd}
	\end{equation}
	cf. diagrams~\ref{eqn:limit dual universal} and \ref{eqn:dual universal}.
\end{rem}

\begin{exmp}[product]\label{exmp:product as universal morphism}
	A product of colour and shape is the universal morphism $(C \times S, \pi)$ shown in the following commutative diagram:
	\begin{equation}\label{eqn:product as universal morphism}
		\begin{tikzcd}
			Z \arrow[d, "u"', dashed] & \Delta(Z) \arrow[rd, "z"] \arrow[d, "\Delta(u)"', dashed]  \\
			C \times S & \Delta(C \times S) \arrow[r, "\pi"'] & (C, S)
		\end{tikzcd}
	\end{equation}
	where $\pi = (\pi_C, \pi_S)$ and $z = (z_C, z_S)$. The objects and arrows in the right side of the diagram are in a category of pair-shaped diagrams, $\mathbf{C}^2$. 
\end{exmp}

\begin{defn}[comma category]\label{defn:comma category}
	Suppose a pair of functors having a common codomain category, $S : \mathbf{A} \rightarrow \mathbf{C}$ and $T : \mathbf{B} \rightarrow \mathbf{C}$. A comma category, written $(S \downarrow T)$, is a category that consists of the following data:
	\begin{itemize}
		\item objects: a triple $(A, B, \gamma)$ for each object $A$ in $\mathbf{A}$, each object $B$ in $\mathbf{B}$ and each arrow $\gamma : S(A) \rightarrow T(B)$ in $\mathbf{C}$,
		
		\item arrows: a pair $(\alpha, \beta)$ for each arrow $\alpha : A \rightarrow A'$ in $\mathbf{A}$ and each arrow $\beta : B \rightarrow B'$ in $\mathbf{B}$ such that diagram
		\begin{equation}
			\begin{tikzcd}
				A \arrow[d, "\alpha"'] & S(A) \arrow[rr, "\gamma"] \arrow[d, "S(\alpha)"'] && T(B) \arrow[d, "T(\beta)"] & B \arrow[d, "\beta"] \\
				A' & S(A') \arrow[rr, "\gamma'"'] && T(B') & B'
			\end{tikzcd}
		\end{equation}
		commutes and
		
		\item composition: the composition of arrow pairs, i.e. $(\alpha', \beta') \circ (\alpha, \beta) = (\alpha' \circ \alpha, \beta' \circ \beta)$, corresponding to the pasting of commutative squares along the shared arrow.
	\end{itemize}
	The identity arrow for each object $(A, B, \gamma)$ is the pair of arrows $(1_A, 1_B)$.
\end{defn}

\begin{rem}\label{rem:universal morphism}
	Universal morphisms are \textit{extremal objects} in comma categories:
	\begin{itemize} 
		\item a universal morphism to object $X$ from functor $F$ is the terminal object in the comma category $(F \downarrow X)$ and		
		
		\item a universal morphism from an object $X$ to a functor $F$ is the initial object in the comma category $(X \downarrow F)$,
	\end{itemize}
	where $X$ is the \textit{constant functor}, i.e. a functor that sends every object and arrow in the domain category to the same object, $X$, and its identity arrow, $1_X$, in the codomain category.
\end{rem}

\begin{exmp}[coproduct]\label{exmp:coproduct-dual-universal}
	Coproduct is dual to product, i.e. a universal morphism from the pair $(A, B)$ to the diagonal functor, which is the initial object in comma category $((A, B) \downarrow \Delta)$.
\end{exmp}

\begin{rem}\label{rem:dualize}
	Other constructions via universal mapping properties likewise dualize. In particular, every limit is a universal cone to a \textit{J}-shaped diagram and the dual construction is \textit{colimit}, which is the universal \textit{cocone} to the same diagram. Moreover, just as every finite limit is obtained from just two kinds of limits, i.e. products and \textit{equalizers}, which are limits to par-shape diagrams---equivalently, from terminal objects and \textit{pullbacks}, which are limits to vee-shape diagrams, every finite colimit is obtained from corresponding duals, i.e. coproducts and \textit{coequalizers}---equivalently, initial objects and \textit{pushouts}.
\end{rem}

\begin{rems}\label{rem:inform}
	Every element $a$ of a set $A$ corresponds to an arrow, written $\overline{a} : * \mapsto a$, also called an \textit{element} or \textit{point} of $A$. This notion of element generalizes to categories other than $\mathbf{Set}$ and elements with shape other than points:
	\begin{itemize}
		\item A (point-shaped) element $f$ of an object $A$ is an arrow $f : 1 \rightarrow A$.
		
		\item An \textit{S}-shaped element $f$ of an object $A$ is an arrow $f : S \rightarrow A$.
	\end{itemize}
	This ``generalized'' notion of an element is just an arrow. Conceptually, however, an arrow $f : S \rightarrow A$ is a causal (informational) relation in the IIT sense of injecting a shape $S$ into an object $A$---cf. a diagram $D : J \rightarrow \mathbf{C}$ as a \textit{J}-shaped generalized element $D$ of a category $\mathbf{C}$.
\end{rems}

\begin{defn}[presheaf]\label{defn:presheaf}
	Suppose a topological space $X$. A \textit{presheaf} is a map $\mathcal{F} : X^\op \rightarrow \mathbf{Set}$ sending each open set $U$ of $X$ to the set $\mathcal{F}(U)$, called the \textit{sections} of $U$, and each inclusion $V \subseteq U$ to the map $\res_{V, U} : \mathcal{F}(U) \rightarrow \mathcal{F}(V)$, called a \textit{restriction morphism}, that satisfies the following conditions:
	\begin{itemize}
		\item $\res_{U, U} : \mathcal{F}(U) \rightarrow \mathcal{F}(U)$ is the identity map for each open set $U$ of $X$ and
		\item if $W \subseteq V \subseteq U$, then $\res_{V, W} \circ \res_{U, V} = \res_{W, U}$ for all open sets $U, V, W$ of $X$.
	\end{itemize}
	The sections of $X$ are called \textit{global sections}.
\end{defn}

\begin{rem}\label{rem:presheaf}
	The conditions for a restriction map are just the identity and compositionality conditions for a functor. Hence, a presheaf is a (set-valued) functor on a topological space, $\mathcal{F} : X^\op \rightarrow \mathbf{Set}$.
\end{rem}

\begin{exmp}[display presheaf]\label{exmp:presheaf}
	Suppose the following sequence of visual display events that starts with a fixation point (``+''):
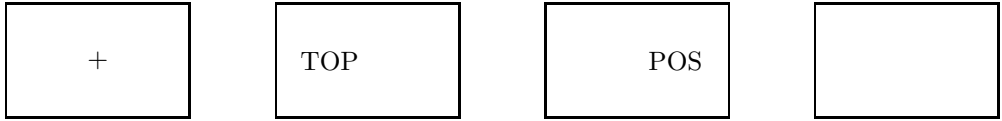
	\begin{equation*}\label{eqn:display}
		\begin{tikzcd}
			\framebox(24mm, 15mm)[c]{+} & \framebox(24mm, 15mm)[c]{\hspace{-1cm}TOP} & \framebox(24mm, 15mm)[c]{\hspace{1cm}POS} & \framebox(24mm, 15mm)[c]{}
		\end{tikzcd}
	\end{equation*}
	For simplicity, the topological space consists of the set $S = \{C, L, R\}$, where $C$ is the centre and $L$ and $R$ are the left and right sides of the visual field (respectively), with the topology $\{\emptyset, \{C\}, \{C, L\}, \{C, R\}, S\}$, which conveys the sense that the centre lies between the left and right fields of view. The display presheaf $\mathcal{D} : S^\op \rightarrow \mathbf{Set}$ has the following assignments for:	
	\begin{itemize}
		\item opensets, $\mathcal{D} : \emptyset \mapsto 1, \{C\} \mapsto \{\mbox{+}\}, \{C, L\} \mapsto \{\mbox{TOP}\}, \{C, R\} \mapsto \{\mbox{POS}\}$, $S \mapsto \emptyset$ and
		\item restrictions, 
		\begin{itemize}
			\item $\mathcal{D} : (\{C, L\} \subseteq S) \mapsto \acute{\pi}, (\{C, R\} \subseteq S) \mapsto \grave{\pi}$, where 
			\begin{itemize}
				\item[] $\acute{\pi} : \emptyset \rightarrow \{\mbox{TOP}\}$ and 
				 $\grave{\pi} : \emptyset \rightarrow \{\mbox{POS}\}$ are the empty (restriction) maps, 
			\end{itemize}
			
			\item $\mathcal{D} : (\{C\} \subseteq \{C, L\}) \mapsto \; !, (\{C\} \subseteq \{C, R\}) \mapsto \; !$, where 
			\begin{itemize}
				\item[] $! : \{\mbox{TOP}\} \rightarrow \{\mbox{+}\}$ and $! : \{\mbox{POS}\} \rightarrow \{\mbox{+}\}$ are the terminal restrictions, and
			\end{itemize}
			
			\item $\mathcal{D} : (\emptyset \subseteq U) \mapsto \; !$, i.e. the restriction maps for the inclusions of the empty set to every open set $U$, are likewise terminal.
		
		\end{itemize}
		
	\end{itemize}
	In words, the presheaf assignments can be interpreted as saying that:
	\begin{itemize}
		\item the fixation point appears on the center of the visual field, ``TOP'' and ``POS'' appear on centre-left and centre-right, and nothing appears on the total visual field, as the stimuli are never presented at the same time, in this sequence of events, and
		
		\item the stimuli appear from out of nowhere in the display sequence.
	\end{itemize}

\end{exmp}

\begin{defn}[open cover]\label{defn:open cover}
	For a topological space $X$, an \textit{open cover} of a set $U$ is a collection of open sets $\{U_i\}_{i \in I}$ of $X$ whose union is $U$.
\end{defn}

\begin{rem}\label{rem:open cover}
	A topology is an open cover of the total set.
\end{rem}

\begin{defn}[sheaf]\label{defn:sheaf}
	A \textit{sheaf} is a presheaf $\mathcal{F} : X^\op \rightarrow \mathbf{Set}$ that satisfies:
	\begin{itemize}
		\item locality (uniqueness): given an open cover $\{U_i\}_{i \in I}$ of an open set $U$, if $s$ and $t$ are sections of $U$, then they restrict to the same section of $U_i$ for all $i \in I$, i.e. $\res_{U_i, U}(s) = \res_{U_i, U}(t)$, and
		
		\item compatibility/gluing (existence): given an open cover $\{U_i\}_{i \in I}$ of an open set $U$, for each pair of sections $s_i$ of $U_i$ and $s_j$ of $U_j$ that restrict to the same section of $U_i \cap U_j$, i.e. $\res_{U_i \cap U_j, U_i}(s_i) = \res_{U_i \cap U_j, U_j}(s_j)$, there exists a section $s$ of $U$ that restricts to $s_i$ and $s_j$, i.e. $\res_{U_i, U}(s) = s_i$ and $\res_{U_j, U}(s) = s_j$.
	\end{itemize}
\end{defn}

\begin{exmp}[display sheaf]\label{exmp:sheaf}
	Suppose the sequence of visual display events (from example~\ref{exmp:presheaf}) is modified to give the appearance of overlapping stimuli (e.g., by placing the stimuli closer to the centre):
	
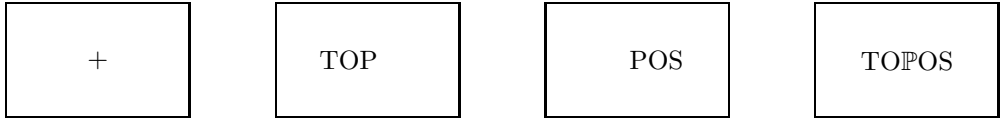
\begin{equation*}\label{eqn:display modified}
		\begin{tikzcd}
			\framebox(24mm, 15mm)[c]{+} & \framebox(24mm, 15mm)[c]{\hspace{-5mm}TOP} & \framebox(24mm, 15mm)[c]{\hspace{5mm}POS} & \framebox(24mm, 15mm)[c]{$\mathrm{TO}\mathbb{P}\mathrm{OS}$}
		\end{tikzcd}
	\end{equation*}
	yielding the following changes in assignments for:	
	\begin{itemize}
		\item opensets, $\{C\} \mapsto \{\mbox{+}, \mathbb{P}\}$, $S \mapsto \{\mathrm{TO}\mathbb{P}\mathrm{OS}\}$ and
		\item restrictions, 
		\begin{itemize}
			\item $\mathcal{D} : (\{C, L\} \subseteq S) \mapsto \acute{\pi}, (\{C, R\} \subseteq S) \mapsto \grave{\pi}$, where 
			\begin{itemize}
				\item[] $\acute{\pi} : \mathrm{TO}\mathbb{P}\mathrm{OS} \mapsto \mbox{TOP}$ and $\grave{\pi} : \mathrm{TO}\mathbb{P}\mathrm{OS} \mapsto \mbox{POS}$, and
			\end{itemize}
			
			\item $\mathcal{D} : (\{C\} \subseteq \{C, L\}) \mapsto \; \acute{p}, (\{C\} \subseteq \{C, R\}) \mapsto \; \grave{p}$, where 
			\begin{itemize}
				\item[] $\acute{p} : \{C, L\} \rightarrow \{C\}$ and $\grave{p} : \{C, R\} \rightarrow \{C\}$ are constant maps picking out $\mathbb{P} \in \{\mbox{+}, \mathbb{P}\}$.
			\end{itemize}
			
		\end{itemize}
		
	\end{itemize}
	This presheaf is a sheaf, as there is just one global section for the only pair of sections, $(\mbox{TOP}, \mbox{POS})$. 
\end{exmp}

\begin{rems}\label{rem:sheaf}
	The unique-existence condition says that a sheaf $\mathcal{F} : X^\op \rightarrow \mathbf{Set}$ satisfies a universal mapping property, hence a sheaf is a universal presheaf. The sheaf condition for sets generalizes to other categories with limits as follows:
	\begin{enumerate}
		\item In a general setting of open sets $U_1$ and $U_2$ and their intersection, $U_1 \cap U_2$, and union, $U = U_1 \cup U_2$, the action of a sheaf $\mathcal{F}$ on open sets and inclusions is shown in the following diagram:
		
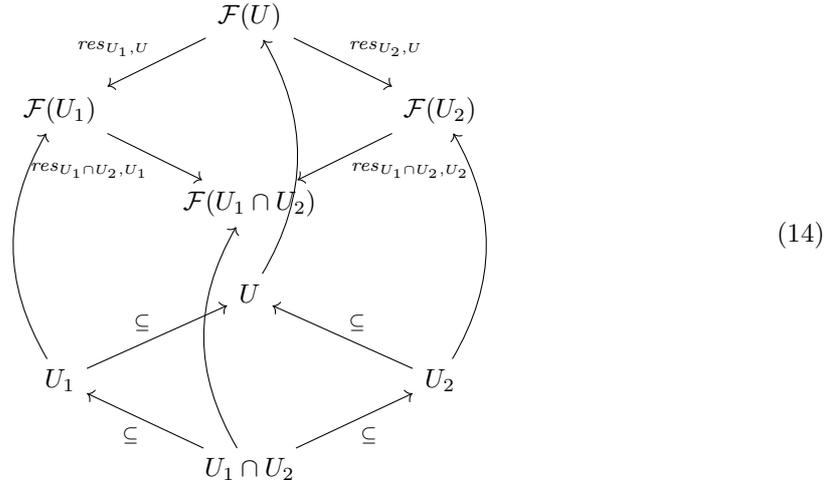
\begin{equation}\label{eqn:sheaf action}
			\begin{tikzcd}
				& \mathcal{F}(U) \arrow[ld, "\res_{U_1, U}"'] \arrow[rd, "\res_{U_2, U}"] \\
				\mathcal{F}(U_1) \arrow[rd, "\res_{U_1 \cap U_2, U_1}"'] && \mathcal{F}(U_2) \arrow[ld, "\res_{U_1 \cap U_2, U_2}"] \\
				& \mathcal{F}(U_1 \cap U_2) \\
				& U \arrow[uuu, bend right] \\
				U_1 \arrow[ru, "\subseteq"] \arrow[uuu, bend left] && U_2 \arrow[lu, "\subseteq"'] \arrow[uuu, bend right] \\
				& U_1 \cap U_2 \arrow[lu, "\subseteq"] \arrow[ru, "\subseteq"'] \arrow[uuu, bend left]
			\end{tikzcd}
		\end{equation}
		The restriction maps for the corresponding inclusions are in the opposite direction, since a sheaf is contravariant functor.
		
		\item The universal mapping property is given as an equalizer of the pair of arrows in the diagram
		\begin{equation}\label{eqn:sheaf}
			\begin{tikzcd}
				\mathcal{F}(U) \arrow[r, ""'] & \prod_i \mathcal{F}(U_i) \arrow[r, ""', shift left] \arrow[r, "", shift right] & \prod_{i, j} \mathcal{F}(U_i \cap U_j)
			\end{tikzcd}
		\end{equation}
		for every open cover $\{U_i\}_{i \in I}$ for every open set $U$ of $X$.
		
		\item The equalizer is equivalently given as a pullback as shown by the following diagram:
		\begin{equation}
			\begin{tikzcd}
				& \mathcal{F}(U) \arrow[ld, "\res_{U_i, U}"'] \arrow[rd, "\res_{U_j, U}"] \arrow[d, "e"'] \\
				\mathcal{F}(U_i) \arrow[rd, "\res_{U_i \cap U_j, U_i}"'] & \mathcal{F}(U_i) \times \mathcal{F}(U_j) \arrow[l, "\pi_1"'] \arrow[r, "\pi_2"] \arrow[d, "\res_1"', shift right] \arrow[d, "\res_2", shift left] & \mathcal{F}(U_j) \arrow[ld, "\res_{U_i \cap U_j, U_j}"] \\
				& \mathcal{F}(U_i \cap U_j)
			\end{tikzcd}
		\end{equation}
		where $e = \langle \res_{U_i, U}, \res_{U_j, U} \rangle$ and $\res_1$and $\res_2$  are given by composition, i.e. $\res_1 = \res_{U_i \cap U_j, U_i} \circ \pi_1$ and $\res_2 = \res_{U_i \cap U_j, U_j} \circ \pi_2$.
		
		\item Limits are terminals in the corresponding comma categories. So, the sections of each open set $U$ of the space $X$ are determined by equalizing the restriction morphisms to the open sets $U_i$ in the cover of $U$.
		
		\item Equivalently, limits are universal morphisms to diagrams. So, the sections are computed as the limits to vee-shaped diagrams picking out the intersections of the open sets $U_i$ in the cover of $U$.
		
	\end{enumerate}	
	So, set-valued sheaves generalize to sheaves valued in any category with finite limits, $\mathbf{C}$, i.e. $\mathcal{F} : X^\op \rightarrow \mathbf{C}$.
\end{rems}

\begin{rem}\label{rem:presheaf-condition}
	A presheaf may satisfy none, one, or both conditions.
	\begin{itemize}
		\item A presheaf that satisfies locality is called a \textit{separated presheaf}.
		\item A presheaf that satisfies compatibility is called a \textit{compatible presheaf}.
		\item A separated compatible presheaf is a sheaf.
	\end{itemize}
\end{rem}

\begin{rem}\label{rem:presheaf-category}
	The collection of presheaves on a topological space $X$ is a category, denoted $\mathbf{Psh}(X)$. The objects are presheaves (functors) and the arrows are presheaf morphisms (natural transformations). Likewise, the collection of sheaves on $X$ is a category, denoted $\mathbf{Sh}(X)$. The category of sheaves is a subcategory of the category of presheaves, $\mathbf{Sh}(X) \subseteq \mathbf{Psh}(X)$.
\end{rem}

\begin{defn}[subobject]\label{defn:subobject}
	Some preliminary definitions are needed: A \textit{monomorphism} is an arrow $i$ such that $i \circ f = i \circ g$ implies $f = g$ for every pair of arrows $f$ and $g$  (cf. set inclusion). Define  an equivalence relation on monomorphisms as follows: a pair of monomorphisms $i : A \rightarrow C$ and $j : B \rightarrow C$ are equivalent, written $i \sim j$, if there exists an isomorphism $k : A \rightarrow B$ such that $i = j \circ k$, as indicated by commutative diagram
	\begin{equation}
		\begin{tikzcd}
			A \arrow[rr, "k", "\cong"'] \arrow[rd, "i"'] && B \arrow[ld, "j"] \\
			& C
		\end{tikzcd}
	\end{equation}
	A \textit{subobject} of an object $C$ is a $\sim$-equivalence class of monomorphisms $[i] = \{j : B \rightarrow C | j \sim i : A \rightarrow C\}$. (Objects and arrows are in a category $\mathbf{C}$.)
\end{defn}

\begin{exmp}[subfunctor]\label{exmp:subfunctor}
	A \textit{subfunctor} is a subobject in a category of functors. So, a \textit{subpresheaf} is a subobject in a category of presheaves; likewise, a \textit{subsheaf} is a subobject in a category of sheaves. For presheaves as tables, one can think of a subpresheaf as a subtable in having a subset of table rows.
\end{exmp}

\begin{rem}\label{rem:subobject}
	The notion of subobject extends to other structures, e.g., a subset is a subobject in the category of sets, a subspace is a subobject in the category of topological spaces, and so on.
\end{rem}

\begin{defn}[subobject classifier]\label{defn:classifier}
	A \textit{subobject classifier} (in a category with finite limits) is an object $\Omega$ together with a monomorphism $t : 1 \rightarrow \Omega$ (from the terminal object) such that for every monomorphism $m : U \rightarrow X$ there exists a unique arrow $\chi_U : X \rightarrow \Omega$ such that 
	\begin{equation}\label{eqn:classifier}
		\begin{tikzcd}
			U \arrow[r, "!"] \arrow[d, "m"'] & 1 \arrow[d, "t"] \\
			X \arrow[r, "\chi_U"'] & \Omega
		\end{tikzcd}
	\end{equation}
	is a pullback diagram. $\Omega$ is called the \textit{classifying object}, also called the \textit{truth object}.
\end{defn}

\begin{exmps}[sets, presheaves]\label{exmp:classifier}
	The categories of sets and presheaves have subobject classifiers:
	\begin{enumerate}
		\item In $\mathbf{Set}$, the classifying object is the Boolean set, $\mathbb{B} = \{0, 1\}$, and the subobject classifier for subsets $U$ of $X$ is given by diagram
		\begin{equation}
			\begin{tikzcd}
				U \arrow[r, "!"] \arrow[d, "\subseteq"'] & 1 \arrow[d, "T"] \\
				X \arrow[r, "\chi_U"'] & \mathbb{B}
			\end{tikzcd}
		\end{equation}
		where $T : * \mapsto 1$ and $\chi_U : x \mapsto 1$ if $x \in U$, otherwise $0$.
		
		\item In a category of presheaves, $\mathbf{Psh}(X)$, the classifying object is the presheaf of open sets $U$ of $X$ and the subobject classifier for a subpresheaf $\mathcal{G}$ of $\mathcal{F}$ acts like ``graded truth'' is returning just those open sets for which $\mathcal{G}$ is ``in'' $\mathcal{F}$. (In diagram~\ref{eqn:classifier}, replace $U$ with $\mathcal{G}$ and $X$ with $\mathcal{F}$, accordingly.) In this situation, the two extremes of empty set and $X$ correspond to false and true, respectively.
	\end{enumerate}
	The collection of subobjects to an object $C$ constitutes a partial order (hence, a category) with $C$ as the terminal object, cf. subspaces of a topological space $X$ with $X$ as the terminal object.
\end{exmps}

\begin{defn}[adjunction]\label{defn:adjunction}
	Given categories $\mathbf{C}$ and $\mathbf{D}$, an \textit{adjunction} $(F, G, \eta, \epsilon, \phi, \psi) : \mathbf{C} \rightharpoonup \mathbf{D}$ consists of:
	\begin{itemize}
		\item a pair of functors $F : \mathbf{C} \rightarrow \mathbf{D}$ and $G : \mathbf{D} \rightarrow \mathbf{C}$, called \textit{left} and \textit{right adjoints}, respectively,
		
		\item a pair of natural transformations $\eta : 1_\mathbf{C} \dotrightarrow G \circ F$ and $\epsilon : F \circ G \dotrightarrow 1_\mathbf{D}$, called \textit{unit} and \textit{counit}, respectively, and
		
		\item a pair of bijections $\phi_{X, Y} : [FX, Y] \rightarrow [X, GY]$ and $\psi_{X, Y} : [X, GY] \rightarrow [FX, Y]$, for each pair of objects $(X, Y)$ in $\mathbf{C} \times \mathbf{D}$, called \textit{left} and \textit{right adjuncts}, respectively, 
	\end{itemize}
	such that there is a universal morphism from $X$ to $G$, i.e.  $(FX, \eta_X)$, for every object $X$ in $\mathbf{C}$ (equivalently, there is a universal morphism from $F$ to $Y$ for every object $Y$ in $\mathbf{D}$). This relationship is called an \textit{adjoint situation}, written $F \dashv G$.
\end{defn}

\begin{exmps}[adjunctions]\label{exmp:adjunction}
	The following are adjoint situations:
	\begin{enumerate}
		\item The diagonal functor is left adjoint to the limit functor, $\Delta \dashv \underset{\longleftarrow}{\Lim}$.
		\item The diagonal functor is right adjoint to the colimit functor, $\underset{\longrightarrow}{\Lim} \dashv \Delta$.
		\item The sheaving functor is left adjoint to the inclusion functor, $\mathit{Sh} \dashv\; \subseteq$, where $\mathit{Sh}$ sends each presheaf to its nearest sheaf by providing just enough sections on each open cover for the given sections on the covered open sets of the presheaf.
		\item A continuous function between topological spaces induces a pair of adjoint functors between the categories of presheaves on those spaces; likewise for categories of sheaves.
	\end{enumerate}
	The sheaf construction involves (co)limits which can be regarded as adding just enough data to complete missing entries (cf. filling out a multiplications table) in the former; dually, as deleting repeated data as a kind of division/quotient (cf. removing repeated table rows) in the latter. These situations may compare to eliciting illusionary feelings and amalgamation of redundant causal relations (see Appendix~B), respectively.
\end{exmps}

\begin{rem}\label{rem:opposing}
	The left and right adjoints are ``dual'' in the sense of being opposed in direction, but the left (right) adjoint is not the corresponding arrow in the opposite category of the category in which the right (left) adjoint resides---both adjoints reside in the same functor category---hence this notion of dual is weaker than, but related to the notion of dual construction (e.g., product and coproduct).
\end{rem}

\begin{rem}\label{rem:adjoints}
	Adjoint functors, $F : \mathbf{C} \rightarrow \mathbf{D}$ and $G : \mathbf{D} \rightarrow \mathbf{C}$, compose as endofunctors, i.e. $G \circ F : \mathbf{C} \rightarrow \mathbf{C}$ and $F \circ G : \mathbf{D} \rightarrow \mathbf{D}$.
\end{rem}

\begin{exmp}[adjointness---order/topology]\label{exmp:adjointness}
	There is an adjunction involving the category of \textit{preordered sets} (i.e. sets with orders that are reflexive and transitive, but not necessarily antisymmetric), denoted $\mathbf{Pre}$, and the category of spaces with \textit{Alexandrov topology} (i.e. arbitrary, not just finite intersections of open sets are open sets), denoted $\mathbf{Alex}$, as follows. 
	\begin{itemize}
		\item Suppose $(P, \le)$ is a preordered set. An \textit{upper set} of $P$ is a set $U \subseteq P$ that is ``upwardly closed'', i.e. for every element $u \in U$ and every element $p \in P$, if $u \le p$ then $p \in U$. In other words, every element above (or following) an element in $U$ is also an element in $U$. The collection of upper sets for $P$ is equivalent to a space with Alexandrov topology. The functor $A : \mathbf{Pre} \rightarrow \mathbf{Alex}$ sends each preordered set to the induced space with Alexandrov topology. 
		
		\item Suppose $(X, T)$ is a topological space. The \textit{specialization preorder} is defined using a ``closure'' operator as follows. First, a \textit{closed set} of $X$ is the complement of an open set in $T$. The \textit{closure} of a set $S \subseteq X$, written $\mathrm{cl}\,S$, is the intersection of all closed sets that contain $S$. A pair of elements $x, y \in X$ has specialization preorder $x \le y$ whenever $x \in \mathrm{cl}\{x\}$. The functor $S : \mathbf{Alex} \rightarrow \mathbf{Pre}$ sends each topological space to its underlying set with the specialization preordered.
		
		\item $A$ is left adjoint to $S$, i.e. $A \dashv S$.
		
	\end{itemize}
	For instance, suppose the order $x \le y \le z$, hence the category with arrows $x \rightarrow y$ and $y \rightarrow z$. The associated space has underlying set $X = \{x, y, z\}$ and Alexandrov topology $\{\emptyset, \{z\}, \{y, z\}, \{x, y, z\}\}$. Going in the other direction, we see that $x$ is in the closure of $y$, i.e.  $x \in \mathrm{cl}\{y\} = \{x, y\}$, hence $x \le y$, and $y$ is in the closure of $z$, i.e. $y \in \mathrm{cl}\{z\} = \{x, y\} \cap \{y, z\} = \{y\}$, hence $y \le z$. If one prefers to reverse the order relations, then the adjunction can be given in terms of ``downwardly closed'' sets, etc. The sense of ``closure'' here is the usual mathematical one: as an operation that adds just enough elements to satisfy some property, e.g., the reflexive closure of a relation $R \subseteq A \times A$ simply adds the pairs $(a, a)$ for each $a \in A$, thus satisfying reflexivity.
\end{exmp}

\newpage

\section{Some conceptual comparisons}

The concepts of axiom, postulate and principle in IIT and category theory (and mathematics, generally) are used in relatable, but not identical senses. Some other comparisons are also given here, particularly as they pertain to the postulates.

\subsection{On axioms, postulates and principles}

``The axioms express properties that are irrefutably true of every conceivable experience---and hence essential properties.'' \cite{grasso2024axiom}. The axiom of existence is ``irrefutable'' in the sense that although one can doubt the truth of the axiom, the act of doubting is itself an experience, therefore experience must exist. Likewise, the other axiomatic properties are irrefutable: e.g., ``every experience is unitary---a whole, irreducible to separate experiences (integration axiom). If we try to refute this by imagining an experience that were not unitary, we end up imagining two or more experiences, each of which is indeed a whole, irreducible to separate experiences.'' \cite{grasso2024glossary}. These axioms are educed from first-person experience, hence the subjective-first view of phenomenology that is characteristic of IIT and supposedly unique to the study of consciousness, in contrast to the objective-first view that prevails in all other sciences \cite{iit2024wiki}. IIT also distinguishes the essential properties (axioms 1--5) from other \textit{accidental properties} that may be found in some, but not all conscious experience, hence are seen as not essential to the existence of consciousness.

Axioms in the mathematical sense are (generally) necessary \textit{and} sufficient conditions for something to be a structure of some kind, although different but equivalent sets of axioms may be used in some situations. For instance, any collection of things and their relations that satisfies the axioms for a category is a category. However, not every category need have a terminal object. If a category has a terminal object, then that object is unique up to a unique isomorphism. So, the axioms for categories correspond to essential properties and the existence/uniqueness of terminal objects to accidental properties in the IIT sense.

The postulates are supposed to express the axioms as properties of physical systems---substrate of consciousness: ``In the IIT method, the postulates are obtained by formulating the essential properties of experience (the axioms) as physical properties of the substrate of consciousness---understood solely in terms of cause-effect power.'' \cite{grasso2024glossary}. The IIT method is a back-and-forth from introspection to formulation that is supposed to establish an identity between phenomenal structure (cf. the extendedness of space) and $\phi$-structure (or, cause-effect structure), i.e. ``the causal distinctions and relations among them'' \cite{hendren2024overview}.

For mathematics, postulate is either a synonym for axiom, or a conjecture (like a hypothesis), i.e. a proposition to be either proved or disproved. 

The IIT principles of maximization and minimization are seen as instances of the category theory principle of determining or defining a structure in terms of a universal mapping property, i.e. a unique-existence condition among all entities (objects) in a context (category) of some kind.

\subsection{On other comparisons}

A basic comparison of other concepts is provided (Table~\ref{table:comparison}) as a guide towards a potentially more detailed category theoretical treatment in future work. For instance, a system (or, substrate) corresponds to a category whose objects and arrows align with the notions of unit and cause-effect relation, respectively; in this way, a self-relation aligns with an endomorphism. However, this comparison is intended to be neither definitive, nor prescriptive, as IIT and category theory concepts often have several aspects from which to view one in terms of the other. In particular, category theory provides several closely related notions of composition, e.g., the composition of arrows $f : A \rightarrow B$ and $g : B \rightarrow C$ as the arrow $g \circ f : A \rightarrow C$, the composition of objects $A$ and $B$, by a UMP, as the product (object) $A \times B$, or constrained product (pullback object) $A \times_C B$, and so on. Furthermore, functors preserve compositions of arrows and (co)limit-preserving functors---which are (left) right adjoints by the \textit{general adjoint functors theorem} \cite<see, e.g.,>{leinster2014basic}---preserve compositions in the forms of products, pullbacks, etc. As has been mentioned elsewhere \cite{phillips2024category}, the classical notion of a \textit{physical instantiation mapping} \cite<see>[]{fodor1988connectionism} between symbols and their physical realizations is essentially a functor. This classical correspondence also suggests comparison with the IIT notion of identifying (extrinsic) physical and (intrinsic) cause-effect structures. If this physical-causal relationship in IIT is also supposed to preserve (co)limits, then necessarily (by the adjoint functors theorem) this relationship is an adjoint situation and so the relevance of this kind of UMP here. For relatively simple situations of phenomenal space, IIT notions of inclusion, connection and fusion align staightforwardly with CT notions of inclusion (subobject), product and coproduct. On the other hand, cause-effect structure as composed of distinctions and relations in IIT also includes other aspects, such as the \textit{cause} and \textit{effect purviews} \cite{grasso2024glossary}, which were not considered in the main text. \cite<See>[for a detailed account of purviews in the context of modelling experimental data.]{leung2021integrated} Some closer comparisons are given in the next section.

\begin{table}[ht]
	\centering
	\begin{tabular}{|l|l|} \hline
		Integrated Information Theory (IIT) & Category Theory (CT) \\ \hline\hline
		system/substrate & category (cf. directed graph) \\ \hline
		unit(s) & object (set) \\ \hline
		cause-effect & arrow ($f : A \rightarrow B$) \\ \hline
		self-relation & endomorphism ($f : A \rightarrow A$) \\ \hline
		inclusion & subobject ($A \subseteq B$) \\ \hline
		composition (connection) & product ($A \times B, \pi$) \\ \hline
		composition---dual (fusion) & coproduct ($A + B, \iota$) \\ \hline
		min-max principles & unique-existence property \\ \hline
		unfolding & $F$-coalgebra (corecursion) \\ \hline
	\end{tabular}
	\caption{A basic comparison of IIT and CT concepts}\label{table:comparison}
\end{table}
 
Another comparison is the IIT concept of unfolding which immediately calls to attention the ``unfold'' operator in functional programming languages, e.g., Haskell \cite{jones2003haskell}, which is based on the category theory concept of an \textit{F-(co)algebra}, i.e. the categorical abstraction and generalization of \textit{(co)recursion}, well-known in computer science \cite<see, e.g.,>[]{bird1997algebra}. These and more elaborate schemes, e.g., \textit{mutual (co)recursion} \cite{hinze2016unifying}, are based on (initial) terminal (co)algebras in corresponding categories, hence the connection to UMPs \cite<see>[for an application to cognition]{phillips2012categorial}.

\subsection{On distinctions and relations}

The two main constituents of a cause-effect structure are distinctions and relations. These concepts are compared with the formal, categorical concept of \textit{span} as a particular kind of (\textit{wedge}-shaped) diagram to a category. A wedge is a category with two diverging arrows, i.e. $-1 \leftarrow 0 \rightarrow 1$, also denoted $\Lambda$. The opposite of a wedge is a vee-shaped diagram, i.e. $-1 \rightarrow 0 \leftarrow 1$, which is used in the definition of a pullback.

\begin{defn}[span]\label{defn:span}
	A span $S$ is a $\Lambda$-shaped diagram to a category $\mathbf{C}$, i.e. a functor $S : \Lambda \rightarrow \mathbf{C}$. 
\end{defn}

\begin{rems}\label{rem:span}
	When a span is given as a pair of arrows, $X \leftarrow Z \rightarrow Y$, the object $Z$ is called the \textit{root}, the objects $X$ and $Y$ are called the \textit{feet}, and the arrows are called the \textit{legs} of the span. A span is also written $S : X \nrightarrow Y$ as there need not be an arrow from $X$ to $Y$ in $\mathbf{C}$. (The definitions of span and cone are similar, but the vertex and legs of a cone are not in the image of a diagram/functor.)
\end{rems}

\begin{exmps}[arrows]\label{defn:arrows}
	Arrows in a category $\mathbf{C}$ are special cases of spans as follows:
	\begin{enumerate}
		\item An arrow $f : A \rightarrow B$ is the span
		\begin{equation}\label{eqn:span arrow}
			\begin{tikzcd}
				& A \arrow[ld, "1_A"'] \arrow[rd, "f"] \\
				A && B				
			\end{tikzcd}
		\end{equation}
		
		\item An arrow $g : B \rightarrow A$ is the span
		\begin{equation}\label{eqn:span arrow opposite}
			\begin{tikzcd}
				& B \arrow[ld, "g"'] \arrow[rd, "1_B"] \\
				A && B				
			\end{tikzcd}
		\end{equation}
		
		\item An identity arrow $1_A$ is the span
		\begin{equation}\label{eqn:span arrow identity}
			\begin{tikzcd}
				& A \arrow[ld, "1_A"'] \arrow[rd, "1_A"] \\
				A && A				
			\end{tikzcd}
		\end{equation}
	\end{enumerate}
\end{exmps}

\begin{exmps}[relations]\label{defn:relations}
	Relations are spans in the category of sets and functions, $\mathbf{Set}$:
	\begin{enumerate}
		\item A product of sets $A$ and $B$ is the span
		\begin{equation}\label{eqn:span product}
			\begin{tikzcd}
				& A \times B \arrow[ld, "\pi_A"'] \arrow[rd, "\pi_B"] \\
				A && B				
			\end{tikzcd}
		\end{equation}
		
		\item The \textit{graph} of a function $f : A \rightarrow B$, i.e. the relation $\Gamma(f) = \{(a, f(a)) | a \in A\}$, is the span
		\begin{equation}\label{eqn:span function}
			\begin{tikzcd}
				& \Gamma(f) \arrow[ld, "\pi_A"'] \arrow[rd, "\pi_B"] \\
				A && B				
			\end{tikzcd}
		\end{equation}
		
		\item A relation $R \subseteq A \times B$ is the span
		\begin{equation}\label{eqn:span relation}
			\begin{tikzcd}
				& R \arrow[ld, "\pi_A"'] \arrow[rd, "\pi_B"] \\
				A && B				
			\end{tikzcd}
		\end{equation}
	\end{enumerate}
\end{exmps}

\begin{defn}[category of spans]\label{defn:category of spans}
	Suppose $\mathbf{C}$ is a category with pullbacks. A \textit{category of spans} on $\mathbf{C}$, written $\mathbf{Span}(\mathbf{C})$, has for:
	\begin{itemize}
		\item objects (0-cells): the objects of $\mathbf{C}$,
		\item arrows (1-cells): the spans
		\begin{equation}\label{eqn:span one cell}
			\begin{tikzcd}
				& Z \arrow[ld, "f"'] \arrow[rd, "g"] \\
				A && B				
			\end{tikzcd}
		\end{equation}
		for each pair of arrows $f : Z \rightarrow A$ and $g : Z \rightarrow B$ in $\mathbf{C}$ and
		\item arrows of arrows (2-cells): the span morphisms $\eta : S \Rightarrow S'$, as indicated by commutative diagram
		\begin{equation}\label{eqn:span two cell}
			\begin{tikzcd}
				& Z \arrow[ldd, "f"', bend right] \arrow[rdd, "g", bend left] \arrow[d, "\eta"'] \\
				& Z' \arrow[ld, "f'"'] \arrow[rd, "g'"] \\
				A && B				
			\end{tikzcd}
		\end{equation}
	\end{itemize}
	where $\eta : Z \rightarrow Z'$ is the component at $Z$, together with composition for:
	\begin{itemize}
		\item 1-cells, denoted $\circ$, defined as pullbacks shown in the following commutative diagram:
		\begin{equation}\label{eqn:span one cell composition}
			\begin{tikzcd}
				&& Z'' \arrow[ld, "\acute{\pi}"'] \arrow[rd, "\grave{\pi}"] \arrow[lldd, "f \circ \acute{\pi}"', bend right] \arrow[rrdd, "k \circ \grave{\pi}", bend left] \\
				& Z \arrow[ld, "f"'] \arrow[rd, "g"] && Z' \arrow[ld, "h"'] \arrow[rd, "k"] \\
				A && B	&& C			
			\end{tikzcd}
		\end{equation}
		i.e. the composition of 1-cells (spans)  $S : A \nrightarrow B$ and $S' : B \nrightarrow C$ is the span $S' \circ S : A \nrightarrow C$, and
		\item 2-cells, denoted $\star$, as the ``pasting'' of spans along the common legs:
		\begin{equation}\label{eqn:span two cell composition}
			\begin{tikzcd}
				& Z \arrow[ldd, "f"', bend right] \arrow[rdd, "g", bend left] \arrow[d, "\eta"'] \\
				& Z' \arrow[ld, "f'"'] \arrow[rd, "g'"] \arrow[d, "\eta'"'] \\
				A & Z'' \arrow[l, "f''"] \arrow[r, "g''"'] & B				
			\end{tikzcd}
		\end{equation}
		i.e. the composition of 2-cells $\eta : S \Rightarrow S'$ and $\eta' : S' \Rightarrow S''$ is the 2-cell $\eta' \star \eta : S \Rightarrow S''$.
	\end{itemize}
\end{defn}

\begin{rem}\label{rem:cospans}
	The dual construction is \textit{cospan}, i.e. a diagram of the form $S : \Lambda^\op \rightarrow \mathbf{C}$.
\end{rem}

By this construction, the relationships between a distinction $D$ and its cause $C$ and effect $E$ purviews may be viewed as a span in $\mathbf{Span}(\mathbf{Set})$ of the form $C \leftarrow D \rightarrow E$. Distinctions compose as 1-cells, i.e. as pullbacks, which affords the chaining of cause-effect relations. A relation $R$ is the irreducible overlap between two distinctions, $D : C \nrightarrow E$ and $D' : C' \nrightarrow E'$, which corresponds to the product $D \times D'$ as the span $R : D \nrightarrow D'$, as shown in the following diagram:
\begin{equation}\label{eqn:relation between distinctions}
	\begin{tikzcd}
		&& R \arrow[ld, "\acute{\pi}"'] \arrow[rd, "\grave{\pi}"]  \\
		& D \arrow[ld, "c"', pos=0.60] \arrow[d, "e", pos=0.45] && D' \arrow[d, "c'"', pos=0.44] \arrow[rd, "e'", pos=0.59] \\
		C & E  && C' & E'
	\end{tikzcd}
\end{equation}

The span construction recovers the phenomenal structure of space as a topology by restricting to spans on the category of sets and inclusions, $\mathbf{Span}(\mathbf{Set}^\subseteq)$: 
\begin{itemize}
	\item inclusion of open sets (spots), $V \subseteq U$, as the span
	\begin{equation}\label{eqn:span inclusion}
		\begin{tikzcd}
			& V \arrow[ld, "1_V"'] \arrow[rd, "\subseteq"] \\
			V && U				
		\end{tikzcd}
	\end{equation}
	\item intersection of open sets (connection of spots), $U \cap V$, as the span
	\begin{equation}\label{eqn:span intersection}
		\begin{tikzcd}
			& U \cap V \arrow[ld, "\subseteq"'] \arrow[rd, "\subseteq"] \\
			U && V				
		\end{tikzcd}
	\end{equation}
	\item and union of open sets (fusion of spots), $U \cup V$, as related by the pullback
	\begin{equation}\label{eqn:span union}
		\begin{tikzcd}
			&& U \cap V \arrow[ld, "\subseteq"'] \arrow[rd, "\subseteq"] \\
			& U \arrow[ld, "1_U"'] \arrow[rd, "\subseteq"] && V \arrow[ld, "\subseteq"'] \arrow[rd, "1_V"] \\
			U && U \cup V	&& V			
		\end{tikzcd}
	\end{equation}
	
\end{itemize}
Hence, by this comparison, distinctions and their relations pertain to UMPs. 

A category of spans has two composition operations and also exemplifies this meta-level approach, as arrows (2-cells) between arrows (1-cells). Such categories are called \textit{2-categories} \cite<see>[]{grandis2020higher}. The second composition is not developed here. However, from a category of spans, $\mathbf{Span}(\mathbf{C})$, one can construct a category ``internal'' to $\mathbf{Span}(\mathbf{C})$ from a particular kind of endofunctor. This construction may be comparable as a categorical analog of the IIT notion of intrinsic causal structure as internal to a physical system as the base category $\mathbf{C}$.

\subsection{On integration and exclusion}

For IIT, the axioms of integration and exclusion as supposed to be independent. Formally, integration and exclusion are closely related, but not identical depending on how the axioms are interpreted. In particular, integration in the mathematical sense, e.g., as used to determine the area under a function, suggests a colimit. As a simple illustration, the ``area'' of two sets $U, V \subseteq X$ given as a Venn diagram is just their union as the coproduct (object) $U + V$ in $\mathbf{Set}^\subseteq$, which automatically excludes all other elements in the Venn diagram, $X$, as the complementary set, $X \setminus U \cap V$. Union is also given by universal morphism from $(U, V)$ to the diagonal function, as shown by commutative diagram
\begin{equation}\label{eqn:venn diagram}
	\begin{tikzcd}
		(U, V) \arrow[r, "\iota"] \arrow[rd, "{(\subseteq, \subseteq)}"'] & \Delta(U \cup V) \arrow[d, "\Delta(\subseteq)", dashed] & U \cup V \arrow[d, "\subseteq", dashed] \\
		& \Delta(X)	& X
	\end{tikzcd}
\end{equation}
where $\iota$ is the inclusions of each set into their union. ($\Delta$ is the \textit{diagonal function} sending each object/arrow to its pair, i.e. $\Delta : U \mapsto (U, U), \subseteq\; \mapsto (\subseteq, \subseteq)$.) So, in terms of a universal morphism, integration in this sense is closely related to exclusion as the mediating arrow relates to the (``complement'' of the) unique arrow. For sets, the complement of $U \subseteq X$ is obtained by negating the classifier for $U$, i.e. $\neg \chi_U = \chi_{X \setminus U}$. The unique arrow is guaranteed to exist for every $X$ (by the UMP), which can be varied independently of $U$ and $V$. So, in this formal sense, the axiom of exclusion is independent of (but closely related to) the axiom of integration. This formal notion of exclusion need not be restricted to sets as \textit{topos theory} generalizes complementation to categories other than $\mathbf{Set}$, not pursued here, such as the category of sheaves (or, presheaves) on a topological space \cite{goldblatt2006topoi, maclane1992sheaves}.

In the IIT sense, however, integration also pertains to not being ``disintegrable'', as in the meaning of ``blackball'' cannot be disintegrated into the meanings of ``black'' and ``ball''. The ``Red sky'' example also illustrated how integration and exclusion are closely related in this sense of being a sheaf vs. being a presheaf that is not a sheaf, as has been proposed for phenomenal experience \cite{youngzie2024towards}. The difference between a presheaf, $\mathcal{F}$, and the ``nearest'' sheaf is given by another UMP in the form of a universal morphism from $\mathcal{F}$ to the inclusion functor $\mathbf{Sh}(X) \subseteq \mathbf{Psh}(X)$, as shown by commutative diagram
\begin{equation}\label{eqn:sheafification}
	\begin{tikzcd}
		\mathcal{F} \arrow[r, "\mathit{sh}"] \arrow[rd, "\phi"'] & \mathcal{F}^+ \arrow[d, "u", dashed] & \mathcal{F}^+ \arrow[d, "u", dashed] \\
		& \mathcal{G}	& \mathcal{G}
	\end{tikzcd}
\end{equation}
The mediating arrow, $\mathit{sh} : \mathcal{F} \rightarrow \mathcal{F}^+$, is called \textit{sheafification}, or \textit{sheaving}. If the presheaf is already a sheaf, then $\mathit{sh}$ is the identity, $1_\mathcal{F}$. Note, however, that this difference is now taken up by the mediating arrow; the unique arrow is the relation from the nearest sheaf, $\mathcal{F}^+$, to each sheaf in $\mathbf{Sh}(X)$. So, the mediating arrow associates with the ``difference'' to the nearest sheaf, however, it does not associate with the difference to other presheaves that are not sheaves. For instance, the two-row table for the ``Red sky'' adage with one row added, say $(R, R)$, is not a sheaf---the $(B, B)$ is still missing---so is not excluded this way. On the other hand, presheaves (including sheaves) are excluded by taking the coproduct of the two subpresheaves constituting the adage by (formal) analogy to the union of two sets (diagram~\ref{eqn:venn diagram}), as the collection of subobjects to an object is a partial order, hence a category.

IIT also excludes \textit{micro-levels} of units from contributing to conscious experience by determining the macro-level units as maximizing integrated information in terms of cause and effect \cite{hoel2016can}. This macro-level vs. micro-level difference suggests \textit{coequalizer} (definition~\ref{defn:coequalizer})---dual to equalizer---with respect to an equivalence relation (example~\ref{exmp:equivalence classes}, remark~\ref{rem:equivalence classes}). A coequalizer is a colimit, hence the connection between the exclusion axiom/postulate and colimit. 

\begin{defn}[coequalizer]\label{defn:coequalizer}
	In a category $\mathbf{C}$, the coequalizer of a pair of arrows $f, g : A \rightarrow B$ is an object $Q$ and an arrow $q : B \rightarrow Q$ such that for every object $Z$ and arrow $z : B \rightarrow Z$ there exists a unique arrow $u : Q \rightarrow Z$ such that the following diagram commutes:
	\begin{equation}\label{eqn:coequalizer}
		\begin{tikzcd}
			A \arrow[rr, "f", shift left] \arrow[rr, "g"', shift right] && B \arrow[r, "q"] \arrow[rd, "z"'] & Q \arrow[d, "u", dashed] \\
			&&& Z			
		\end{tikzcd}
	\end{equation}
	$Q$ is called the \textit{quotient object}.
\end{defn}

\begin{exmp}[equivalence classes]\label{exmp:equivalence classes}
	In $\mathbf{Set}$, suppose a set $A$ and an equivalence relation $R$ on $A$, i.e. a set $R \subseteq A \times A$ and two projections $\pi_1, \pi_2 : R \rightarrow A$. The coequalizer of $\pi_1$ and $\pi_2$ is the smallest set of equivalence classes, written $A/R$, and the assignment of the elements in $A$ to their equivalence classes, as shown by commutative diagram	
	\begin{equation}\label{eqn:equivalence}
		\begin{tikzcd}
			R \arrow[rr, "\pi_1", shift left] \arrow[rr, "\pi_2"', shift right] && A \arrow[r, "q_R"] & A/R 
		\end{tikzcd}
	\end{equation}
	$A/R$ is called the \textit{quotient set}.
\end{exmp}

\begin{rem}\label{rem:equivalence classes}
	Every surjective (onto) function $f : A \rightarrow B$ induces a set of equivalence classes on the domain, $A$. Elements $a$ and $a'$ in $A$ are $f$-equivalent, written $a \sim_f a'$, when they map to the same element $f(a) = f(a')$ in $B$. An $f$-equivalence class for an element $a \in A$ is the set $[a]_f = \{a' \in A | f(a') = f(a)\}$. The quotient set is $A/f = \{[a]_f | a \in A\}$.
	
\end{rem}

This situation compares with IIT in the sense that two units $a$ and $a'$ with the same effect on an element $b$ are amalgamated as the same distinction. An important aspect of this step is in regard to grain: ever finer physical distinctions do not necessarily contribute to consciousness (so, say, molecules need not be conscious), because such smaller ``atoms'' of the physical system may all act (i.e. take and make a difference) in the same way---the macro-level of interaction can override the micro-level of interaction \cite{hoel2016can}. The coequalizer expresses this situation as a UMP.

\end{document}